\documentclass[a4paper,11pt]{article}

\pdfoutput=1
\usepackage[british]{babel}
\usepackage{jcappub}
\usepackage{graphicx}
\usepackage{array}
\usepackage{tabularx}
\usepackage{float}
\usepackage{xcolor}
\usepackage{physics}
\usepackage{slashed}
\usepackage[font=small, labelfont=bf]{caption}
\usepackage[labelformat=simple]{subcaption}

\usepackage{amsmath, amsthm, amssymb, amsfonts}
\usepackage{multirow}

\definecolor{myblue}{rgb}{0.3, 0.5, 0.7}
\definecolor{myyellow}{rgb}{0.85, 0.72, 0.37}

\def\beq{\begin{align}}
\def\eeq{\end{align}}
\newcommand{\bi}{\begin{itemize}}
\newcommand{\ei}{\end{itemize}}
\newcommand{\ben}{\begin{enumerate}}
\newcommand{\een}{\end{enumerate}}
\newcommand{\be}{\begin{equation}}
\newcommand{\ee}{\end{equation}}
\newcommand{\bea}{\begin{eqnarray}}
\newcommand{\eea}{\end{eqnarray}}

\newcommand{\V}{\mathcal{V}}

\renewcommand{\O}{\mathcal{O}}

\newcommand{\de}{\partial}

\renewcommand{\i}{{\rm in}}

\begin{document}

\title{Gravitational Waves from Multiple Cosmic Superstrings and the Overshoot Problem}
\author[a,b]{Luca Brunelli,}
\author[a,b]{Michele Cicoli,} 
\author[a,c]{Muhammad Hassan,}
\author[a]{Seyed Ehsan Qoreishi,}
\author[a,b]{Francisco G. Pedro}
\affiliation[a]{Dipartimento di Fisica e Astronomia, Universit\`a di Bologna, via Irnerio 46, 40126 Bologna, Italy}
\affiliation[b]{INFN, Sezione di Bologna, viale Berti Pichat 6/2, 40127 Bologna, Italy}
\affiliation[c]{Fakultät Physik, Technische Universität Dortmund, D-44221 Dortmund, Germany}

\emailAdd{l.brunelli@unibo.it}
\emailAdd{michele.cicoli@unibo.it}
\emailAdd{muhammad.hassan115@studio.unibo.it}
\emailAdd{seyedehsan.qoreishi@studio.unibo.it}
\emailAdd{francisco.soares@unibo.it}

\abstract{  
Post inflationary string cosmology can feature an initial population of multiple species of cosmic superstrings whose tension is controlled by a modulus rolling over a steep potential toward a late-time minimum. We perform a full analysis of the associated dynamical system, finding that overshooting the minimum is prevented by the friction of a radiation background of gravitational waves produced from the early decay of effective strings arising from NS5- and D3-branes wrapped around internal cycles. On the other hand, fundamental strings survive longer and decay when the modulus is oscillating around the minimum and they have about $1/3$ of the total energy density. The spectrum of gravitational waves generated by the decays of these multiple cosmic superstrings, even if diluted by a late epoch of modulus domination, can still result in a high-frequency, multi-peaked signal, offering an observational signature of generic features of string theory. 
}

\maketitle 

\section{Introduction}

Two generic features of string compactifications, which can arise in a large class of cosmological models, are fundamental strings (F-strings) and moduli rolling over steep potentials for several Planck units (see  \cite{Cicoli:2023opf} for a comprehensive review on string cosmology). Given that in string theory all mass scales are moduli-dependent, the tension of F-strings varies with time when the moduli evolve. 

This observation has recently triggered various studies on the evolution of cosmic strings with time-dependent tension \cite{Conlon:2024uob, Revello:2024gwa, Brunelli:2025ems, Ghoshal:2025tlk, SanchezGonzalez:2025uco, Brunelli:2025eif, Conlon:2025mqt, Brunelli:2025dif, Chun:2025ret, Brunelli:2026qkp}. Most of the attention has been focused on type IIB LVS models \cite{Balasubramanian:2005zx, Conlon:2005ki,Cicoli:2008va} where the underlying dynamical system involves kinating moduli, radiation-like background fluids and different species of cosmic strings, including F-strings and effective strings arising from NS5-branes wrapped around internal $4$-cycles (NS5-strings) and D3-branes wrapping $2$-cycles (D3-strings). 

In particular, ref. \cite{SanchezGonzalez:2025uco} focused on F-strings and a modulus rolling over an exponential potential, finding an attractor where the total energy density is dominated by the cosmic superstrings. This behaviour has a benefit and a drawback. On the one hand, a large amount of energy into F-strings after inflation can lead to a sizeable gravitational wave (GW) signal at high frequencies, even if diluted by the late time decay of the modulus \cite{Conlon:2025mqt}. On the other hand, given that the potential energy density of the modulus goes to zero in the F-string loop tracker, the modulus would have enough energy to climb the barrier that separates its late time minimum from the decompactification limit. This is the famous overshoot problem \cite{Brustein:1992nk}. This issue can be avoided by advocating the presence of a radiation background fluid acting as a Hubble friction \cite{Kaloper:1991mq, Barreiro:1998aj, Brustein:2004jp, Kaloper:2004yj, Battefeld:2005av,Itzhaki:2007nk, Conlon:2008cj, Acharya:2008bk}, but in these scenarios the UV origin of a sufficiently large radiation component is unclear \cite{Conlon:2022pnx,Mosny:2025cyd}.

Subsequently, ref. \cite{Brunelli:2025eif} extended the analysis in two ways: ($i$) by considering the whole LVS potential instead of just its exponential approximation which is valid only in the region far from the minimum; ($ii$) by studying the evolution of different kinds of cosmic strings (NS5- and D3-strings, in addition to F-strings), albeit one at a time. The results of \cite{Brunelli:2025eif} revealed that the case with NS5-strings does not feature any overshoot problem even in the absence of radiation, due to the presence of an attractor fixed point where the energy density of the Universe is dominated by NS5-strings which absorb energy from the rolling modulus very efficiently. 

In this paper, we improve the analysis of this complicated cosmological system by including all possible ingredients: all three types of cosmic superstrings at the same time (NS5-, D3- and F-strings), a modulus evolving over the whole LVS potential, and the background radiation arising due the production of GWs from the decay of each species of cosmic superstrings. Being heavier, NS5- and D3-strings decay into GWs very quickly, sourcing a background of radiation which dominates the energy density of the Universe. The Hubble friction on the modulus dynamics caused by this radiation fluid allows it to avoid overshooting. Being lighter, F-string survive for longer and decay into GWs while the modulus is oscillating around the minimum of its potential. The resulting present-day spectrum of stochastic GWs features a multi-peaked signal at high frequencies which can pave the way for the detection of generic features of string theory in the sky. 

This paper is organised as follows. In Sec. \ref{sec:NoEmission} we analyse in detail the dynamics of the autonomous system involving three species of cosmic superstrings and a kinating modulus, paying particular attention to the solution of the overshoot problem with and without background radiation. In Sec. \ref{sec:GW Emission} we then deepen our study by including the effect of the GW emission from the decay of each species of cosmic superstrings, computing at the end the current GW spectrum at high frequency. We discuss our conclusions in Sec. \ref{Concl}, while App. \ref{app:GW tensor} explains why perturbation theory is still valid even when the total energy density is almost all in GWs.

\section{Multiple cosmic superstrings and the overshoot problem}
\label{sec:NoEmission}

\subsection{Multiple cosmic superstrings with a time-varying tension}

The system we aim to study comprises a scalar field $\phi$ subject to a potential $V$ in a flat FLRW background:
\begin{equation}
    ds^2 = -dt^2 + a^2(t) \, d \boldsymbol{x}^2\,, 
\end{equation} 
and different species of cosmic superstring loops coupled to the field through their tension which turns out to be exponential in $\phi$:
\begin{equation}
\label{eq:exponential tension}
    \mu(\phi) = \mu_0 \, e^{-\sqrt{6}\, \beta\,\phi/M_p}\,.
\end{equation}
In our setting, what distinguishes the different string species is simply the value of $\beta$. The cosmic string loops behave as a cosmological fluid, whose energy density depends on their tension and number density $n_{\rm loop}$, as well as on their physical length $\ell$. 
In fact, by definition, the energy of a loop is given by the product of its tension by its length:
\begin{equation}
E_{\rm loop} = \ell \, \mu\,.
\label{Eloop}
\end{equation}
The loops behave as a fluid with energy density:
\begin{equation}
\label{eq:rho_loop}
\rho_{\rm loop}(t) = E_{\rm loop} (t) \, n_{\rm loop}(t) = \ell(t) \, \mu(t) \, n_{\rm loop}(t)\,. 
\end{equation}
For the time being, we shall neglect the loss in energy density due to GW emission, which will then be introduced in Sec. \ref{sec:GW Emission}. 

This means that the only way loops gain or lose energy is through the variation of their tension, which for small loops, directly translates into a variation of length \cite{Conlon:2024uob, Revello:2024gwa}:
\begin{equation}
\label{eq:l(t) no GW}
    \ell (t) = \ell_{\rm in} \sqrt{\frac{\mu_{\rm in}}{\mu(t)}}\,.
\end{equation}
Throughout this work, we will assume the loops are non-interacting, so that their number density simply redshifts as $n_{\rm loop}(t) \sim a^{-3}$. Therefore:
\begin{equation}
\label{eq: rho_loop no GW}
    \rho_{\rm loop} (t) =  \rho_{\rm loop}^\i \left(\frac{a_\i}{a(t)}\right)^{3} \sqrt{\frac{\mu(t)}{\mu_\i}}\,,
\end{equation}
where $\rho_{\rm loop}^\i = \ell_\i\, \mu_\i \, n_{\rm loop}^\i$, and the index `in' indicates the initial value of the quantity. While the $a^{-3}$ redshift is common to all loop species, each redshifts differently due to a different time-dependence of $\mu(t)$, inherited from the time evolution of $\phi(t)$ through the exponential coupling in \eqref{eq:exponential tension}. Let us now describe the dynamical system we will study in the first part of this work. 

\subsection{The dynamical system}

Let us consider a scalar field $\phi(t)$ subject to a potential $V(\phi)$ in a FLRW background. Moreover, we assume the presence of a perfect, barotropic background fluid with equation of state $\rho_{\rm f} = \omega \, p_{\rm f}$, and $N_s$ different species of isolated, non-interacting string loop fluids. Each of these string fluids is characterised by a tension of the form \eqref{eq:exponential tension} with a different $\beta_j$. With this content, the Friedman constraint reads:
\begin{equation}
\label{eq:Friedman constraint}
    3 H^2 M_p^2= \frac{1}{2} \dot \phi^2 + V(\phi) + \rho_{\rm f} + \sum_{j=1}^{N_s} \rho_j  \equiv \rho_{\rm tot}\,, 
\end{equation}
where $\rho_j$ is the energy density of the $j$-th string loop fluid. The continuity equation for the scalar field reduces to the Klein-Gordon equation in an expanding Universe:
\begin{equation}
\label{eq:KG equation}
    \ddot \phi + 3H \dot \phi + V_{, \phi} + \sum_{j=1}^{N_s} \rho_{j,\phi} = 0\,,
\end{equation}
while for the background fluid and each loop species, we have:
\begin{eqnarray}
\dot \rho_{\rm f} &=& -3 H (\omega + 1) \rho_{\rm f}\,, \\
\dot \rho_j &=& -3 H \rho_j + \rho_{j,\phi} \,  \dot  \phi\,.
\end{eqnarray}
To analyse the behaviour of the system, it is useful to introduce some dimensionless phase space dynamical variables, which represent the fraction of energy density of the Universe present at each time in each component. These are:
\begin{align}
    X^2 & \equiv \frac{\dot \phi^2/2}{ \rho_{\rm tot}}\equiv \Omega_{\rm k}\,, \label{eq:X}\\
Y^2 & \equiv \frac{V}{ \rho_{\rm tot}}\equiv \Omega_V\,, \label{eq:Y}\\
Z_j^2 & \equiv \frac{\rho_j}{ \rho_{\rm tot}} \equiv \Omega_{j}\,, \label{eq:Z}\\
W^2 & \equiv \frac{\rho_{\rm f}}{ \rho_{\rm tot}}\equiv \Omega_{\rm f}\,. \label{eq:W} 
\end{align}
We shall also define the total energy density fraction in loops as:
\begin{equation}
\label{eq:Omega loop}
    \Omega_{\rm loop} = \sum_{j = 1}^{N_s} \Omega_j\,.
\end{equation}
Using these definitions, the Friedman constraint \eqref{eq:Friedman constraint} reduces to:
\begin{equation}
\label{eq:FC in phase space}
    X^2 + Y^2 + \sum_{j =1}^{N_s} Z_j^2 + W^2= 1\,.
\end{equation}
Differentiating the definitions \eqref{eq:X}-\eqref{eq:W} with respect to time and using the second Friedman equation:
\begin{equation}
\label{eq:second Friedman}
    \dot H  = -\frac{1}{2 M_p^2}\left[\dot \phi^2 + \sum_{j = 1}^{N_s}\rho_{j} + (\omega+1)\rho_{\rm f}\right],
\end{equation}
we can get an autonomous system in phase space. To do so, let us define the potential parameter:
\begin{equation}
\lambda(\phi) \equiv -\sqrt{\frac{2}{3}}\frac{V_{,\phi}}{V}\,,
\end{equation}
and trade the time variable $t$ for the number of e-foldings $N = \ln a$, so that $' \equiv d/dN$. The equations in phase space then reduce to:
\begin{align}
X' &= \frac{3}{2}\bigg[X\left(X^2-Y^2+\omega W^2-1\right) 
+ \lambda(\phi)\, Y^2+ \sum_j \beta_j Z_j^2 \bigg], \label{eq:X'}\\
Y' &= \frac{3}{2}Y\left[X^2-Y^2+\omega W^2-\lambda(\phi) X+1\right], \label{eq:Y'}\\
Z_j' &= \frac{3}{2}Z_j\left[X^2-Y^2+\omega W^2-\beta_j X\right], \label{eq:Z_i'}\\
W' &= \frac{3}{2}W\left[X^2-Y^2+\omega \left(W^2-1\right)\right], \label{eq:W'} 
\end{align}
where we also used the exponential dependence of the tension in \eqref{eq:exponential tension}. This dynamical system features $N_s+3$ equations of which only $N_s+2$ are independent. In fact, when numerically solving the system of differential equations, one of the $Z_j$'s can be eliminated by using \eqref{eq:FC in phase space}. The fixed points of this system are very similar to those of the dynamical system in \cite{SanchezGonzalez:2025uco, Brunelli:2025eif}, with the difference that there are $N_s$ cases of those fixed points involving loops. To give an explicit example, we show in Tab. \ref{tab:fixed points no GW} the full classification of fixed points for the case $N_s=3$, indicating the $\beta_j$'s as $\beta_F$, $\beta_3$ and $\beta_5$ for later convenience.  

\begin{table*} 
\centering
\resizebox{\textwidth}{!}{
\begin{tabular*}{1.431\textwidth}{|c|c|c|c|c|c|c|c|}
\hline
\centering \textbf{FP} & \textbf{X} & \textbf{Y} & $\mathbf{Z_F}$ & $\mathbf{Z_3}$& $\mathbf{Z_5}$& $\mathbf{W}$ &  \textbf{Existence} \\
\hline
$\mathcal{K}$ & 1 & 0 & 0 & 0 & 0 & 0 & $\forall  \lambda\,$ and $\,\forall \beta_j$ \\
\hline
$\mathcal{M}$  & $\frac{\lambda}{2}$ &  $\sqrt{1-\frac{\lambda^2}{4}}$ &  0  & 0  & 0  & 0  &$\forall \beta_i,\, \lambda \leq 2$ \\
 \hline
 $\mathcal{L}_F$ &  $\beta_F$ & 0 & $\sqrt{1-\beta_F^2}$   &  0 & 0& 0  &$\forall \lambda, \, \beta_F  \leq 1$  \\
 \hline 
 $\mathcal{L}_3$ &  $\beta_3$ & 0 & 0 & $\sqrt{1-\beta_3^2}$& 0& 0  &$\forall \lambda, \, \beta_3  \leq 1$  \\
 \hline 
$\mathcal{L}_5$ &  $\beta_5$ & 0 & 0 & 0& $\sqrt{1-\beta_5^2}$& 0  &$\forall \lambda, \, \beta_5  \leq 1$  \\
\hline
$\mathcal{F}$ & 0& 0 &0 &0 &0 &1 & $\forall  \lambda\,$ and $\,\forall \beta_j$\\
\hline
$\mathcal{T}_{1}^{(F)}$ & $\frac{1}{\lambda - \beta_F}$ &   $\frac{\sqrt{\beta_F^2 +1 - \lambda\beta_F }}{\lambda-\beta_F}$ & $\frac{\sqrt{\lambda^2 -2 - \lambda\beta_F }}{\lambda-\beta_F}$ & 0 & 0 & 0 & $\beta_F \leq 1 \;$ and $\; \frac{\beta_F}{2} + \sqrt{\frac{\beta_F^2}{4}+2} \leq \lambda \leq \beta_F + \frac{1}{\beta_F}$\\
\hline
$\mathcal{T}_{1}^{(3)}$ & $\frac{1}{\lambda - \beta_3}$ &   $\frac{\sqrt{\beta_3^2 +1 - \lambda\beta_3}}{\lambda-\beta_3}$ & 0& $\frac{\sqrt{\lambda^2 -2 - \lambda\beta_3 }}{\lambda-\beta_3}$  & 0 & 0 & $\beta_3 \leq 1 \;$ and $\; \frac{\beta_3}{2} + \sqrt{\frac{\beta_3^2}{4}+2} \leq \lambda \leq \beta_3 + \frac{1}{\beta_3}$\\
\hline
$\mathcal{T}_{1}^{(5)}$ & $\frac{1}{\lambda - \beta_5}$ &   $\frac{\sqrt{\beta_5^2 +1 - \lambda\beta_5}}{\lambda-\beta_5}$ & 0& 0 & $\frac{\sqrt{\lambda^2 -2 - \lambda\beta_5 }}{\lambda-\beta_5}$  & 0 & $\beta_5 \leq 1 \;$ and $\; \frac{\beta_5}{2} + \sqrt{\frac{\beta_5^2}{4}+2} \leq \lambda \leq \beta_5+ \frac{1}{\beta_5}$\\
\hline 
$\mathcal {S}$ &         $\frac{\omega+1}{\lambda}$&         $\frac{\sqrt{1-\omega^2}}{\lambda}$& 0 & 0& 0& $\sqrt{1-\frac{2(\omega+1)}{\lambda^2}}$& $\lambda \geq \sqrt{2(\omega+1)}\,$, $\, \beta_j \neq \lambda \frac{\omega}{(\omega+1)}$  \\
\hline
$\mathcal{T}_2^{(F)}$  &      $\frac{\omega}{\beta_F}$ & 0& $\frac{\sqrt{\omega(1-\omega)}}{\beta_F}$&0 &0 &$\sqrt{1-\frac{\omega}{\beta_F^2}}$ & $ \forall \lambda,\, \beta_F \geq \sqrt{\omega}$ \\
\hline
$\mathcal{T}_2^{(3)}$  &      $\frac{\omega}{\beta_3}$ & 0& 0&$\frac{\sqrt{\omega(1-\omega)}}{\beta_3}$ &0 &$\sqrt{1-\frac{\omega}{\beta_3^2}}$ & $ \forall \lambda,\, \beta_3 \geq \sqrt{\omega}$ \\
\hline
$\mathcal{T}_2^{(5)}$  &      $\frac{\omega}{\beta_5}$ & 0& 0& 0 &$\frac{\sqrt{\omega(1-\omega)}}{\beta_5}$ &$\sqrt{1-\frac{\omega}{\beta_5^2}}$ & $ \forall \lambda,\, \beta_5 \geq \sqrt{\omega}$ \\
\hline
\end{tabular*}
}
\caption{Fixed points (FP) of the autonomous system \eqref{eq:X'}-\eqref{eq:W'} with existence conditions.}
\label{tab:fixed points no GW}
\end{table*}

\subsection{Type IIB embedding}

\begin{figure}[h!]
\centering
\includegraphics[width=0.65\textwidth]{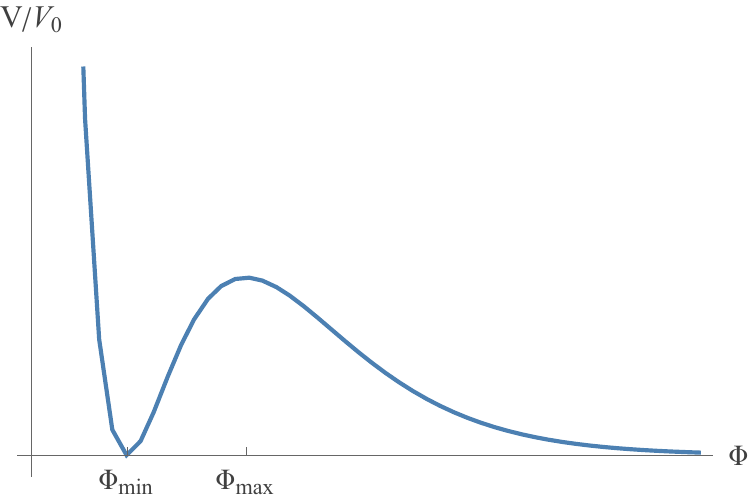}
\caption{LVS potential \eqref{eq:V_LVS} for the canonically normalised volume $\Phi$ for $\epsilon = 0.013$ and $\delta = 5.397 \times10^{-12}$.}
\label{fig:LVS potential}
\end{figure}

Let us consider the embedding of this system in type IIB string compactifications. We will identify the scalar field $\phi$ with the canonically normalised volume modulus of the internal Calabi-Yau manifold:
 \begin{equation}
 \label{eq:Phi}
     \frac{\Phi}{M_p} = \sqrt{\frac{2}{3}}\, \ln \V\,.
 \end{equation}
We focus on the case where this field is subject to a potential $V$ with a Minkowski or dS minimum at $\Phi_{\rm min}$. For $\Phi < \Phi_{\rm min}$, the potential exhibits a steep, approximately exponential behaviour, whereas to the right of the minimum it features a comparatively shallow maximum at $\Phi_{\rm max}$. A well-motivated realisation of this example is the LVS potential \cite{Balasubramanian:2005zx, Conlon:2005ki,Cicoli:2008va}, whose minimum is obtained by balancing higher derivative corrections against non-perturbative effects. The vacuum energy can be ensured to be non-negative by adding a contribution to $V$ from an appropriate hidden sector (e.g. an $\overline{D3}$-brane \cite{Kachru:2003aw}, a T-brane \cite{Cicoli:2013cha,Cicoli:2015ylx} or non-zero F-terms of the complex structure moduli \cite{Saltman:2004sn,Gallego:2017dvd,Hebecker:2025tui}). The result is a potential of the form:
\begin{equation}
\label{eq:V_LVS}
    V_{\rm LVS}(\Phi) = V_0 \left[(1- \epsilon \, \Phi^{3/2}) \,e^{- 3 \sqrt{{\frac{3}{2}}}\, \Phi} + \delta \, e^{-2 \sqrt{\frac{3}{2} \Phi}} \right],
\end{equation}
where $\epsilon$ and $\delta$ depend on the details of the compactification, and are treated here as tunable parameters. In particular, we choose their values so that the minimum of the potential has vanishing energy, $V(\Phi_{\rm min}) = 0$, at $\Phi_{\min} = 19\,M_p$, which is approximately the largest value the volume can take while avoiding the cosmological moduli problem (CMP) \cite{Coughlan:1983ci, Banks:1993en, deCarlos:1993wie}. This potential is depicted in Fig. \ref{fig:LVS potential}. 

Typical inflationary models where the volume modulus acquires a time-dependence are brane-antibrane inflation \cite{Burgess:2001fx, Dvali:2001fw,Kachru:2003sx,Baumann:2007ah,Baumann:2007np,Cicoli:2024bwq}  where $\mathcal{V}$ behaves as a waterfall field at the end of inflaton, and volume modulus inflation \cite{Conlon:2008cj,Cicoli:2015wja} where $\mathcal{V}$ plays the role of the inflaton at an inflection-point in the region at low $\Phi$. The volume modulus (be it the inflaton or a spectator during inflation) undergoes a large field excursion $\Delta \Phi \sim \O(10)\,M_p$ from where it sat during inflation and its late-time minimum $\Phi_{\rm min}$. The potential displayed in Fig. \ref{fig:LVS potential} is exponentially steep to the left of the minimum, approximately behaving as $V \simeq V_0\, e ^ {- 3\,  \sqrt{3/2}\, \Phi}$. On the other hand, the maximum located at $\Phi_{\rm max}$ is comparatively shallow. This can give rise to the so-called \textit{overshoot problem} \cite{Brustein:1992nk, Kaloper:1991mq, Barreiro:1998aj, Brustein:2004jp, Kaloper:2004yj, Battefeld:2005av, Conlon:2008cj, Acharya:2008bk, Conlon:2022pnx,Mosny:2025cyd}, where the field does not slow down sufficiently to stabilise at its minimum, but instead rolls past it, climbs up the maximum and rolls down towards infinity. To avoid this problem, a mechanism is needed that is sufficiently efficient at draining kinetic energy from the rolling field, thus effectively braking its descent. 
Ref. \cite{Brunelli:2025eif} considered the consequence that a single string loop fluid has on a field rolling down the potential \eqref{eq:V_LVS} towards its late-time minimum. They found that only string species with low values of $\beta$ ($\beta <0.23$) help the field avoid overshooting.

In type IIB, there are three species of strings whose tension depends on the volume in a different way. We have F-strings, whose tension is set simply by the string scale $M_s$ which, in 4D Einstein frame, is:
\begin{equation}
\label{eq:mu_f}
\mu_{F} \simeq M_s^2 = \frac{\sqrt{g_s}\, M_p^2}{4 \pi \V}\,.
\end{equation}
Upon canonical normalisation of the volume, we see that:
\begin{equation}
\mu_F \simeq M_p^2 \, e^{-\sqrt{\frac{3}{2}} \Phi}\,,
\end{equation}
so that, comparing with \eqref{eq:exponential tension}, $\beta_F = 1/2$. Alternatively, we could have an \textit{effective string}, i.e. the result in the EFT of a $p$-brane wrapping a $(p-1)$-cycle. In type IIB CY compactifications with O3/O7-planes, the only BPS-stable states that can be built in this way are \textit{D3-strings} (D3-branes wrapping 2-cycles) and \textit{NS5-strings} (NS5-branes wrapping 4-cycles).\footnote{One could in principle also have D7-strings, where a D7-brane wraps the whole 6D internal manifold, but, as shown in \eqref{eq:effective string tension}, this effective string would not depend on $\mathcal{V}$, i.e. $\beta= 0$, and so it is not interesting for our purposes.} As shown in \cite{Brunelli:2025ems}, the tension of these strings acquires a dependence on the volume of the internal cycle $\Sigma_{p-1}$ wrapped by the brane:
\begin{equation}
\label{eq:effective string tension}
    \mu_p \simeq M_s^{p+1} \, \text{Vol}(\Sigma_{p-1})\,.
\end{equation}
If we assume that the brane wraps the cycle corresponding to the volume modulus, then by \eqref{eq:effective string tension} we get:
\begin{equation}
\label{eq:D3 tension}
    \mu_3 \simeq M_s^2 \, \V^{1/3} \simeq M_p^2 \,e^{-\frac{2}{3}\sqrt{\frac{3}{2}} \Phi}\,,
\end{equation}
and:
\begin{equation}
\label{eq:NS5 tension}
\mu_5 \simeq M_s^2 \, \V^{2/3} \simeq M_p^2\, e^{-\frac{1}{3}\sqrt{\frac{3}{2}} \Phi}\,,
\end{equation}
meaning that $\beta_3 = 1/3$ and $\beta_5 = 1/6$.\footnote{D1-strings (usually referred to as \textit{D-strings}) and $(p,q)$-strings (bound states of $p$ F-strings and $q$ D-strings) are also BPS-stable states. However, the dependence of their tension on $\mathcal{V}$ is the same as \eqref{eq:mu_f}, and so their treatment is analogous to that of F-strings for the purpose of this work. The difference lies in the dependence of their tension on the dilaton, which could lead to interesting dynamics when combined with the evolution of the volume. This is left as a direction for future work.}

The main result of \cite{Brunelli:2025eif} was that a sufficient initial concentration of NS5-strings  can drain enough kinetic energy from the rolling volume to avoid overshooting, since $\beta_5=1/6<0.23$.  The first question we want to answer in this work is: what changes when all three loop species are present at the same time? The analysis follows from the dynamical system \eqref{eq:X'}-\eqref{eq:W'} using similar techniques as those developed in \cite{Brunelli:2025eif}.

\subsection{The overshoot problem without background radiation}

We first study the stability of the system against overshooting with no background radiation. The dynamical system is obtained from \eqref{eq:X'}-\eqref{eq:W'} simply by setting $W = 0$.

\begin{figure}[h!]
\centering
\includegraphics[width=0.65\textwidth]{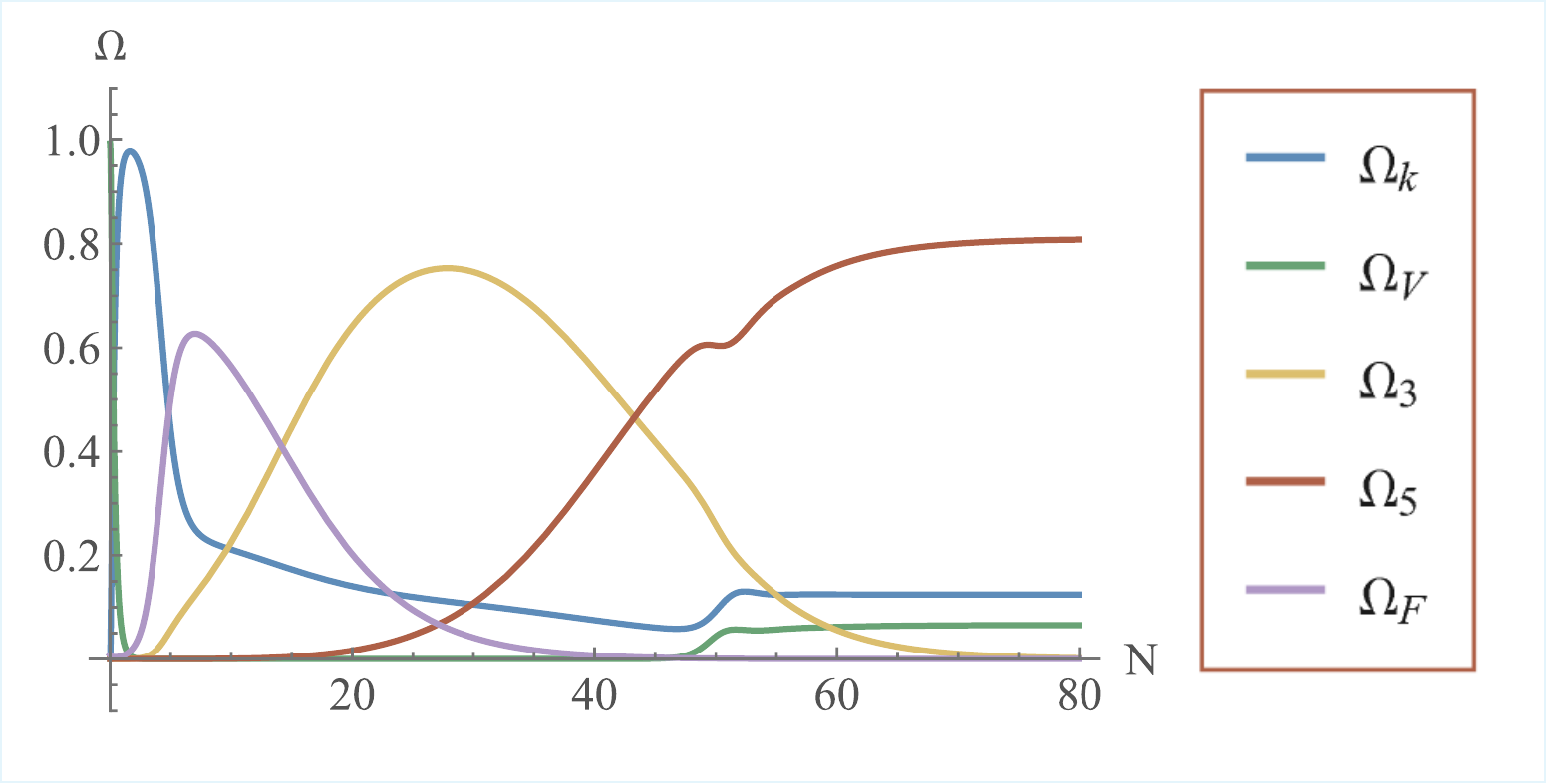}
\caption{Evolution of the energy densities of the system with an exponential potential with $\lambda =3$. The initial conditions are chosen to get a clear distinction between the three loop-dominated eras, setting $\Omega_k^{\i} = 0$, $\Omega_F^{\i} = 0.8 \times 10^{-2}$, $\Omega_3^{\i} = 10^{-4}$ and $\Omega_5^{\i} = 10^{-8}$.}
    \label{fig:multiple loops exponential no fluid}
\end{figure}

We first analyse what happens in the case of a pure exponential potential with $\lambda = 3$, which is a good approximation of \eqref{eq:V_LVS} for $\Phi< \Phi_{\rm min}$. From \eqref{eq:exponential tension} and \eqref{eq: rho_loop no GW}, we can see that:
\begin{equation}
\label{eq:rho_j redshift}
    \rho_j \sim a^{-3} \V^{-\beta_j}\,,
\end{equation}
and so, as the volume is running down its potential, a lower value of $\beta_j$ indicates a slower redshift. In \cite{Brunelli:2025eif}, it was argued that such different redshifts of the energy density of different loop fluids would lead the NS5-strings, whose dependence on the volume is the weakest, to dominate the energy density. Given enough time, this is indeed what happens, as can be seen in Fig. \ref{fig:multiple loops exponential no fluid}.

Here one can appreciate that, if the initial conditions are chosen appropriately, the system undergoes three separate phases of string-loop domination: first by F-strings, then by D3-strings and finally by NS5-strings, with the fixed point $\mathcal{T}^{(5)}_{1}$ as final attractor, in which $\Omega_5^{(\mathcal{T}_1^{(5)})} \simeq  0.809$ as can be seen in Tab. \ref{tab:fixed points no GW}. Before reaching the final attractor, the system first moves towards $\mathcal{L}_F$. However, the energy density in D3-strings also grows and quickly drives the system towards $\mathcal L_3$ before it can reach $\mathcal L_F$. While the system is approaching $\mathcal L_3$, however, NS5-strings start being relevant, and the system finally moves towards $\mathcal{T}_1^{(5)}$. 

Let us now consider the full potential \eqref{eq:V_LVS}. The main difference from the purely exponential case is that the field excursion between the initial value $\Phi_\i$ and the value of $\Phi_{\rm min}$ is finite. 
The first problem we would like to address is whether we can get stabilisation in a scenario in which the initial energy density in each loop species is comparable. 

\begin{figure}[h]
\centering
\includegraphics[width=0.65\textwidth]{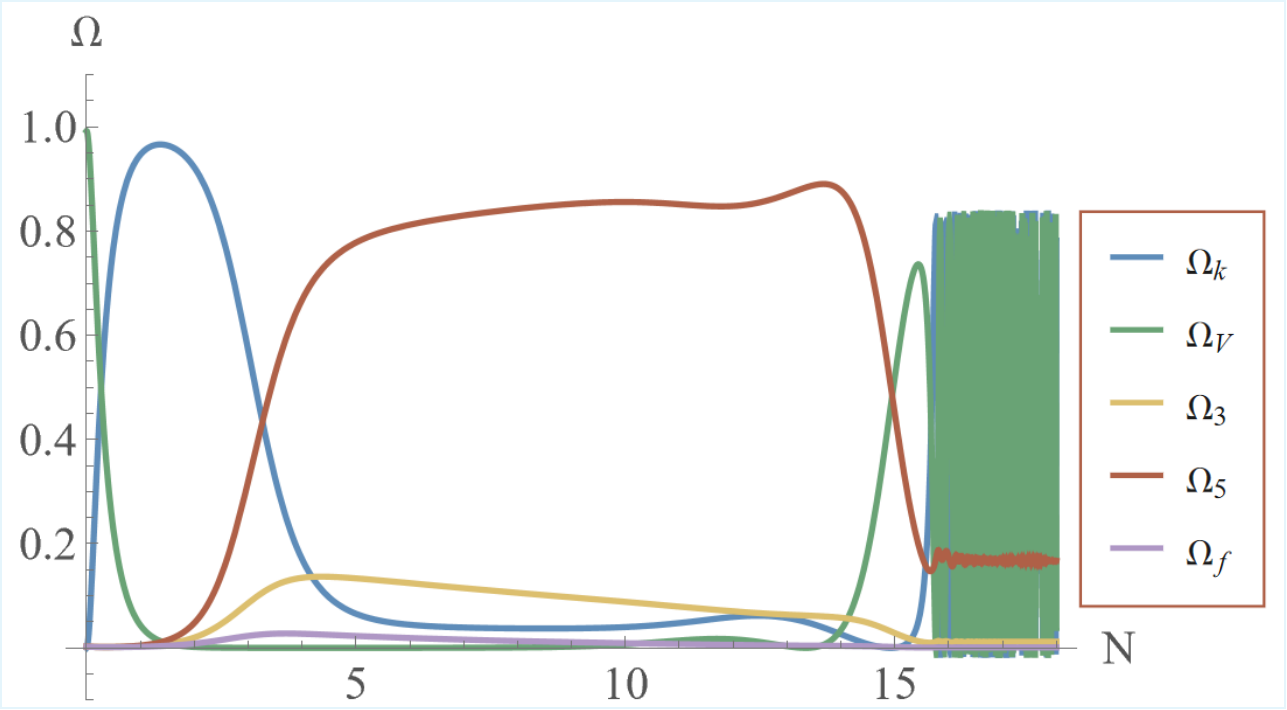}
\caption{Evolution of the energy densities of the system with $\Omega_{\rm loop}^{\i}=0.008$, $\Omega_F = \Omega_3 = \Omega_{5} = \Omega_{\rm loop}/3$, $\Phi_\i = 6 M_p$ and $\dot \Phi_\i = 0$. Note the oscillations of $\Phi$ around its minimum.}
    \label{fig:energy density no fluid equi}
\end{figure}

The answer depends on the initial total amount of energy density fraction in loops. In particular, in the case of equipartition of initial energy density among species (i.e. $\Omega_j^{\i} = \Omega_{\rm loop}^{\i}/3$), and initial conditions for the field of $\Phi_\i= 6 M_p$ (i.e. initial volume of $\V_\i = e^{\sqrt{3/2}\Phi_\i} \simeq 1554$, large enough to have an EFT under control) and $\dot \Phi_\i=0$, we get no overshooting for $\Omega_{\rm loop}^{\i} \gtrsim 0.0043$. The evolution of the energy density fractions is displayed in Fig. \ref{fig:energy density no fluid equi} for $\Omega_{\rm loop}^{\i} \simeq 0.008$.

As can be appreciated therein, a major role in the stabilisation is played by NS5-strings, as was expected. Given that the initial energy densities for each loop species are the same, the NS5-strings, which redshift the slower, come to dominate the energy density of the Universe as soon as the system moves away from kination. 

\begin{figure}[h]
\centering
\includegraphics[width=0.65\textwidth]{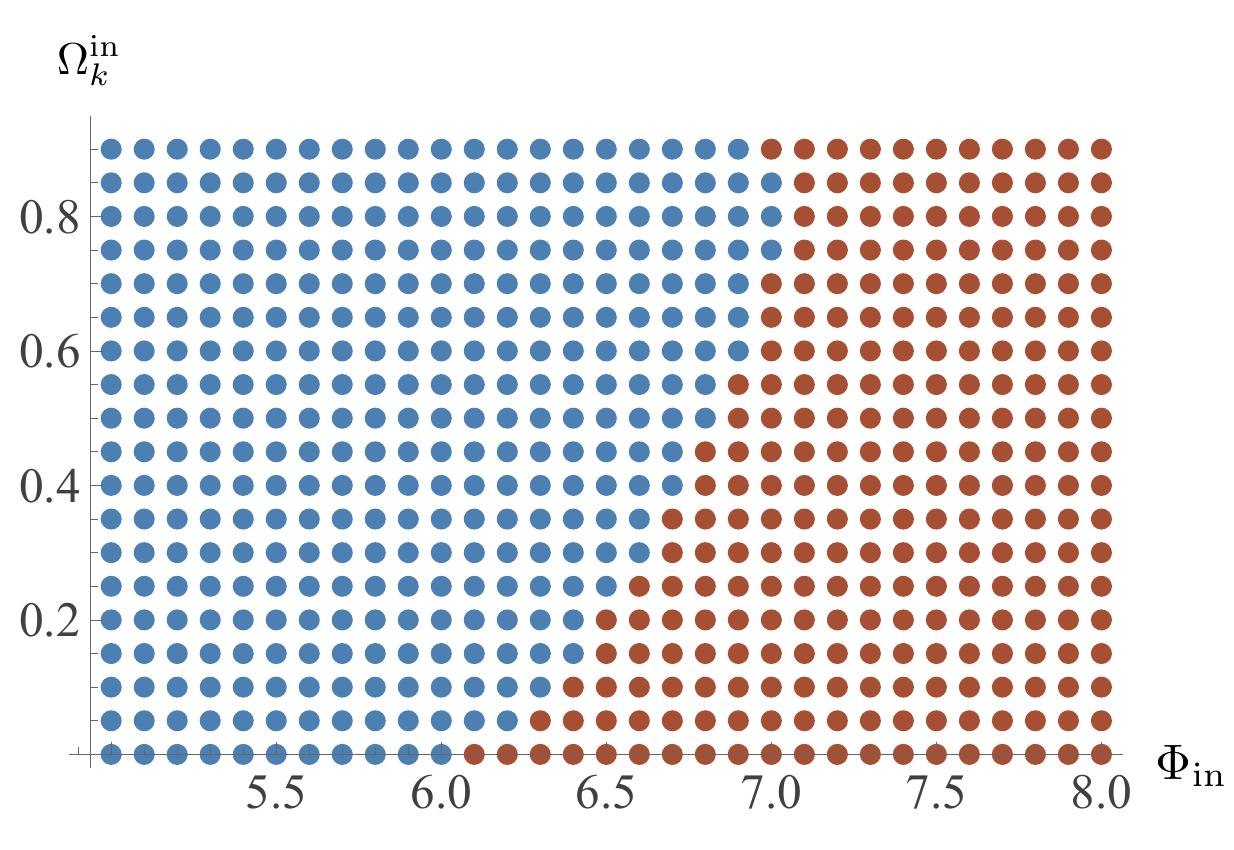}
\caption{Scan of initial conditions for $\Phi_\i$ and $\Omega_k^{\i}$ with $\Omega^{\i}_{\rm loop}=4.5 \times 10^{-3}$. Blue dots correspond to stabilisation, while red dots to overshooting.}
\label{fig:X_Phi Scan No Fluid low loops}
\end{figure}

To better understand the behaviour and stability of the system, we performed various scans on the initial conditions. First, we explored what happens when the field has a non-zero initial velocity. The result of a first scan is shown in Fig. \ref{fig:X_Phi Scan No Fluid low loops}.

Note that in this simple case (at least for $\Omega_k^{\i}< 0.9$) the maximal initial value of $\Phi_{\rm in}$ for which there is no overshooting seems to be an increasing function of $\Omega_k^{\i}$.
This behaviour is expected and is observed also in the case of a single NS5-string fluid. The reason this happens is because with a larger initial velocity, the system starts closer to kination, which, as shown in Fig \ref{fig:energy density no fluid equi}, tends to be the first fixed point the system visits. In kination, the field redshifts as $\rho_{\Phi} \sim a^{-6}$, while the string fluids redshift as in \eqref{eq:rho_j redshift}. Since in kination the volume grows as $\V \sim t \sim a^{3}$ \cite{Conlon:2024uob}, the overall redshift is $\rho_j \sim a^{-3(1+ \beta_j)}>a^{-6}$ $\forall \beta_j < 1$. On the other hand, when $\dot \Phi_\i = 0$, the system spends a few efoldings transitioning between a potential-domination state where $\rho \sim \text{const}$, and kination. During this period, the relative energy density in loops does not grow, and so their braking effects kicks in later. 

\begin{figure}[h]
\centering
\includegraphics[width=\textwidth, height= 0.3 \textwidth]{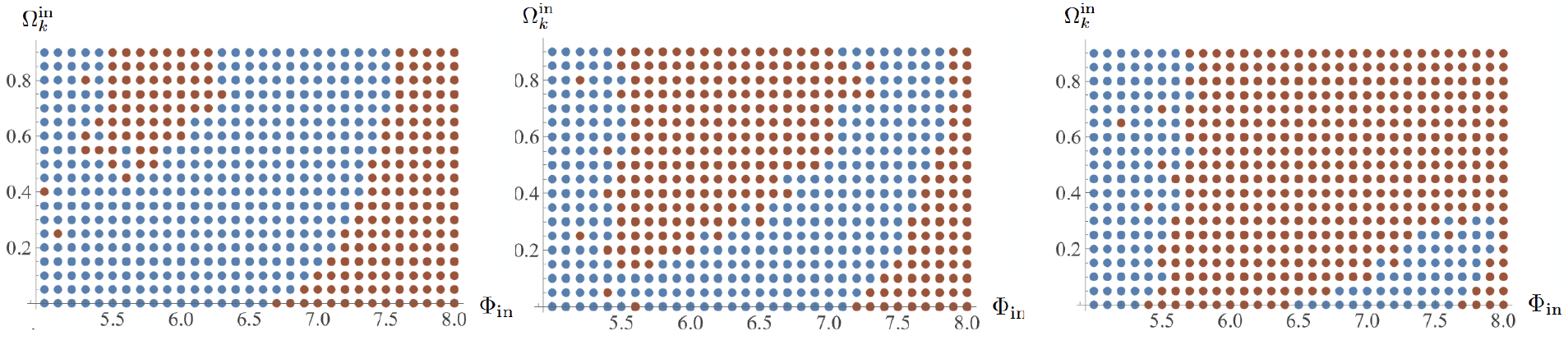}
\caption{Scan of initial conditions for $\Phi_\i$ and $\Omega_k^{\i}$ for various values of $\Omega^{\i}_{\rm loop}$. In each case $\Omega_j^{\i} = \Omega^{\i}_{\rm loop}/3$. Blue dots correspond to stabilisation, while red dots to overshooting. Left to right: $\Omega_{\rm loop}^{\i}=0.008\,, 0.012\,, 0.02$.}
\label{fig:X Phi various Omegaloop}
\end{figure}

Things get trickier when we increase $\Omega_{\rm loop}^{\i}$, as shown in Fig. \ref{fig:X Phi various Omegaloop}. The behaviour displayed here is peculiar to the case of multiple loop species and does not apply to the case of NS5-strings only. In fact, in the case of just NS5-strings, the field would never overshoot in the last case displayed in Fig. \ref{fig:X Phi various Omegaloop}. 

The same effect we noticed in Fig. \ref{fig:X_Phi Scan No Fluid low loops} is present in Fig. \ref{fig:X Phi various Omegaloop} as well; in fact, in the latter,  the right-most branch of the blue area indicating stabilisation follows a similar curve as the one in the former. Moreover, by increasing the total fraction of energy density in loops, we are also increasing $\Omega_5^{\i}$, which is good to avoid overshooting. This shows in Fig. \ref{fig:X Phi various Omegaloop} as the right branch of the blue area is pushed towards larger values of $\Phi_\i$. What instead is peculiar to these plots is the growing red dent starting from high $\Omega_k^{\i}$ and eating its way down, effectively cutting the stabilisation regions in two disconnected areas as $\Omega_{\rm loop}^{\i}$ increases. This behaviour is due to the increasing fraction of energy density in D3-strings. As argued in \cite{Brunelli:2025eif}, D3-strings do not avoid overshooting. However, their energy density redshifts slower than F-strings, making them play a more significant role in the system's evolution. This can also be appreciated in Fig. \ref{fig:energy density no fluid equi}, where we can see that D3-strings hold a large fraction of the energy density throughout the system's evolution ($0.1 \lesssim \Omega_3\lesssim 0.2$ between $N=3$ and $N=15$). 

As a matter of fact, loops are coupled to the modulus and can exchange energy with it, draining energy from the field, but also giving it back. In fact, the term $\sum_j \beta_j Z_j^2$ in \eqref{eq:X'} is always positive. Larger $\beta_j$'s hence contribute to a larger acceleration of the field. This is why D3-strings and F-strings not only do not help in solving the overshoot problem but make it worse through this energy injection.
In the case at hand, the relevant species causing destabilisation are D3-strings, since $Z_3^2$ always corresponds to a sizeable fraction of the energy density. In light of this effect, we shall analyse Fig. \ref{fig:X Phi various Omegaloop} in terms of the fixed points of the dynamical system.

The left vertical blue branch, present in all three panels of Fig. \ref{fig:X Phi various Omegaloop}, is due to the braking effect of NS5-strings in the final attractor $\mathcal T_{1}^{(5)}$. Since the field starts at lower initial values, the field reaches its attractor further from the minimum, and NS5-strings have more time to brake its race, hence resulting in a more likely stabilisation. The blue branch on the right, on the other hand, is due to the braking of the field by means of NS5-strings in the saddle point $\mathcal{L}_5$. What happens in between the two branches is a peculiar competition between the redshifts of D3-strings and NS5-strings. When the system moves between $\mathcal L_5$ and $\mathcal{T}_1^{(5)}$, it undergoes a slight acceleration, also visible in the plot in Fig. \ref{fig:energy density no fluid equi} towards $N = 13$, where $\Omega_k$ increases. This is due to the fact that $X_{\mathcal{L}_5} < X_{\mathcal{T}_1^{(5)}}$ as visible from Tab. \ref{tab:fixed points no GW} with $\beta_5 = 1/6$. This acceleration makes the field gain back some kinetic energy from the loops, and if it is not followed by a sufficiently long period of braking, may undermine the stabilisation. This acceleration is not peculiar to the multi-fluid system, as it is present in the case of a single loop fluid as well. However, the presence of other string species, in particular D3-strings, makes this acceleration period last longer, as the system has a larger phase space to move in. This makes it more unstable and prone to overshoot. 

\begin{figure}
\centering
\includegraphics[width=0.65\textwidth]{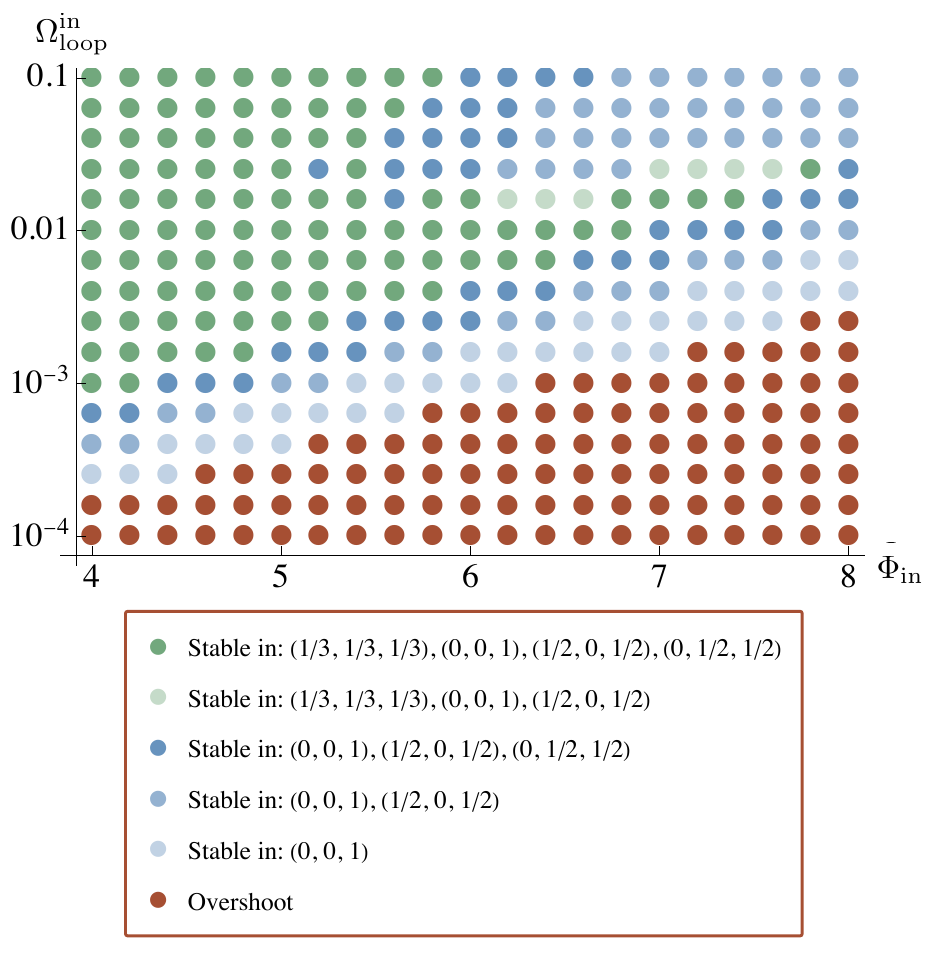}
\caption{Stabilisation map of the field in the absence of radiation with multiple loop species. The vertical axis is on a logarithmic scale and in each case we set $\Omega_k^{\i} = 0$. Different colours indicate different fractions of energy density in each loop species $\left(\Omega_F^{\i}, \Omega_3^{\i}, \Omega_5^{\i}\right)/\Omega_{\rm loop}^{\i}$.}
 \label{fig:loops vs phi_0 no rad}
\end{figure}

We performed a further scan over the initial energy density in loops and the initial condition for the field. Here we are varying both $\Omega_{\rm loop}^{\i}$ and the relative fractions belonging to each loop species. The result of the scan is displayed in Fig. \ref{fig:loops vs phi_0 no rad}. What can be clearly appreciated is that the most-stable case is when only NS5-strings are present (light-blue area), which was expected. Moreover, the upper-left corner of the figure is stable in every configuration where some NS5-strings are present. This is due to the large fraction of energy in loops which directly translates into a large fraction of NS5-strings. Moreover, on the left we are at small field values, indicating a longer braking period. As we  move right we need a larger relative fraction of NS5-strings to stabilise the field due to the shorter field range. A rare example of \textit{constructive} interference among loop species is the green horizontal stripe at $\Omega_{\rm loop}^{\i} \sim 0.01$. Here the same competing effects among the redshifts of D3- and NS5-strings allow the field to have a prolonged braking, which improves stabilisation. 

\subsection{The overshoot problem with background radiation}

\begin{figure}[h]
\centering
\includegraphics[width=0.65\textwidth]{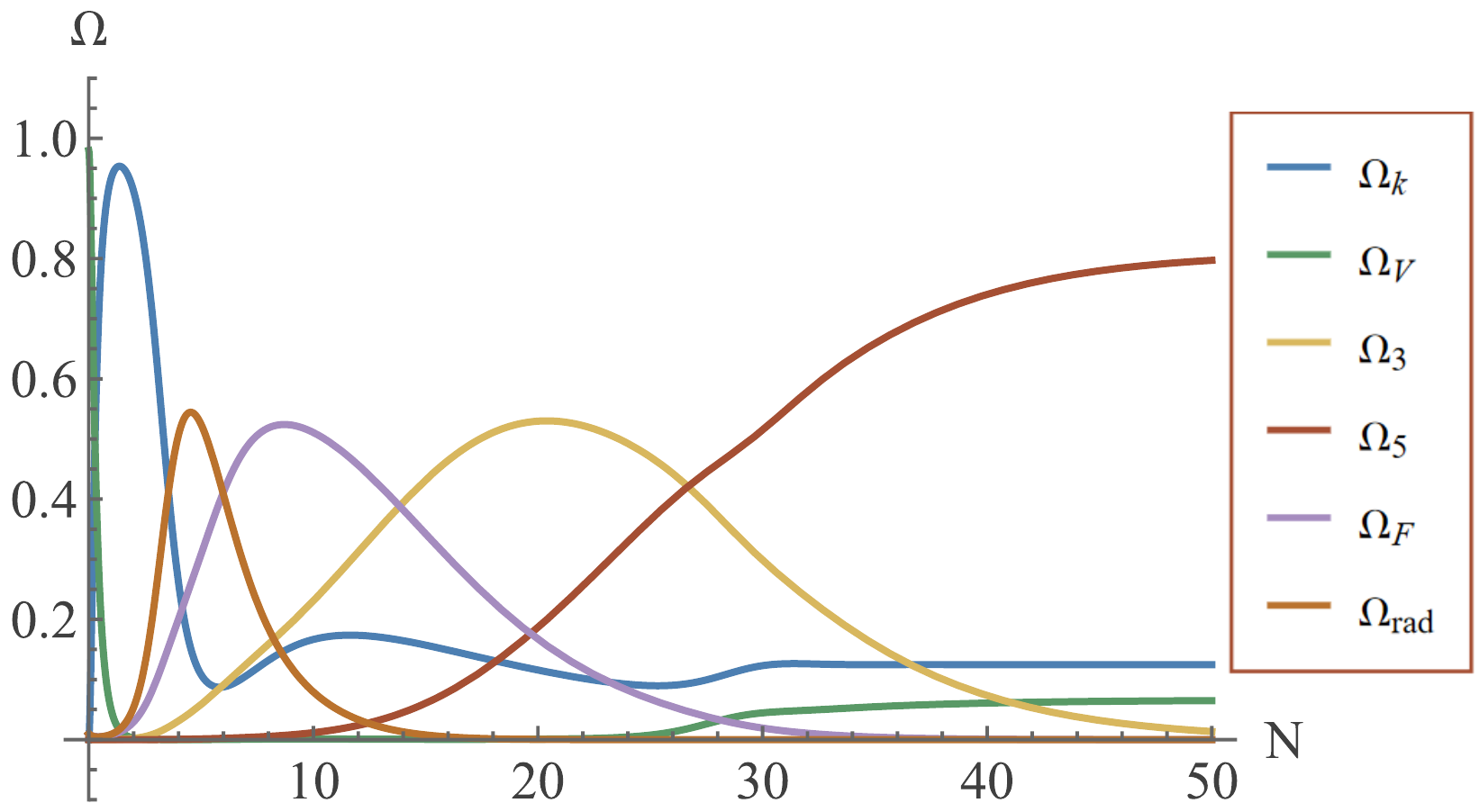}
\caption{Evolution of the energy densities of the dynamical system \eqref{eq:X'}-\eqref{eq:W'} with radiation and an exponential potential with $\lambda =3$. The initial conditions are $\Omega_k^{\i}=0$, $\Omega_F^{\i}= 10^{-2}$, $\Omega_3^{\i}= 3 \times 10^{-4}$, $\Omega_5^{\i} = 10^{-6}$ and $\Omega_{\rm rad}^{\i}=10^{-2}$, chosen to show different eras in the evolution.}
    \label{fig:multiple loops exponential radiation}
\end{figure}

We now turn to study the full autonomous system \eqref{eq:X'}-\eqref{eq:W'}, including a background fluid. By adding the fluid to the system, we are effectively enlarging its parameter space, since we now also have to consider the initial fraction of energy density in the fluid, $\Omega_{\rm f}^{\i}$. For the sake of simplicity, and to make contact with what we will discuss later, we consider a radiation background fluid, setting $\omega = 1/3$. The evolution of the energy densities in the case of an exponential potential is shown in Fig. \ref{fig:multiple loops exponential radiation}. 

\begin{figure}[h]
    \centering
\includegraphics[width=0.65\textwidth]{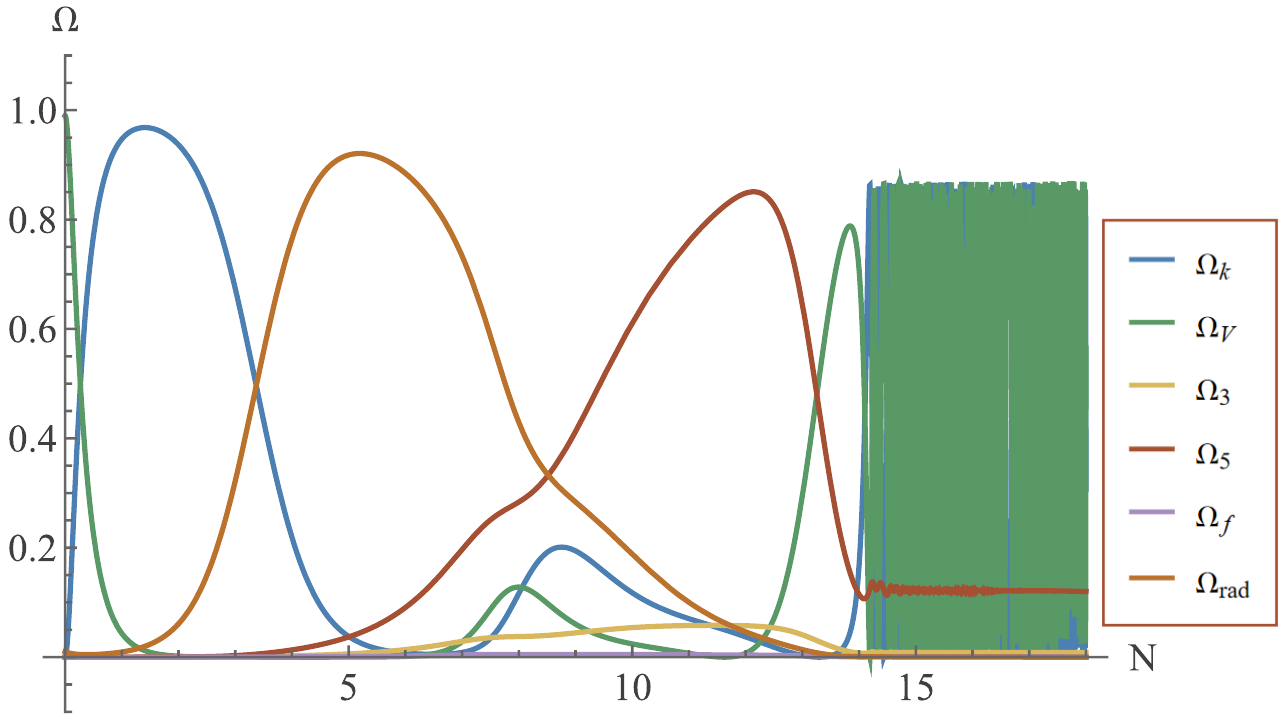}
\caption{Evolution of the energy densities in the presence of radiation. The initial conditions are $\Phi = 6 M_p$, $\Omega_k^{\rm in}=0$, $\Omega_{\rm rad}^{\rm in}= 10^{-2}$, and $\Omega_{\rm loop}^{\rm in} = 5 \times 10^{-5}$, with equal repartition among loop species. The field overshoots without radiation.}
\label{fig:energy density with radiation}
\end{figure}

The final attractor is once again $\mathcal T_1^{(5)}$, but the system explores various saddle points before reaching it. Stabilisation is possible and aided in some regimes by the additional friction due to radiation. In Fig. \ref{fig:energy density with radiation} we show an exemplar case, in which the field was overshooting the minimum without radiation.

\begin{figure}[h]
\centering
\includegraphics[width=0.65\textwidth]{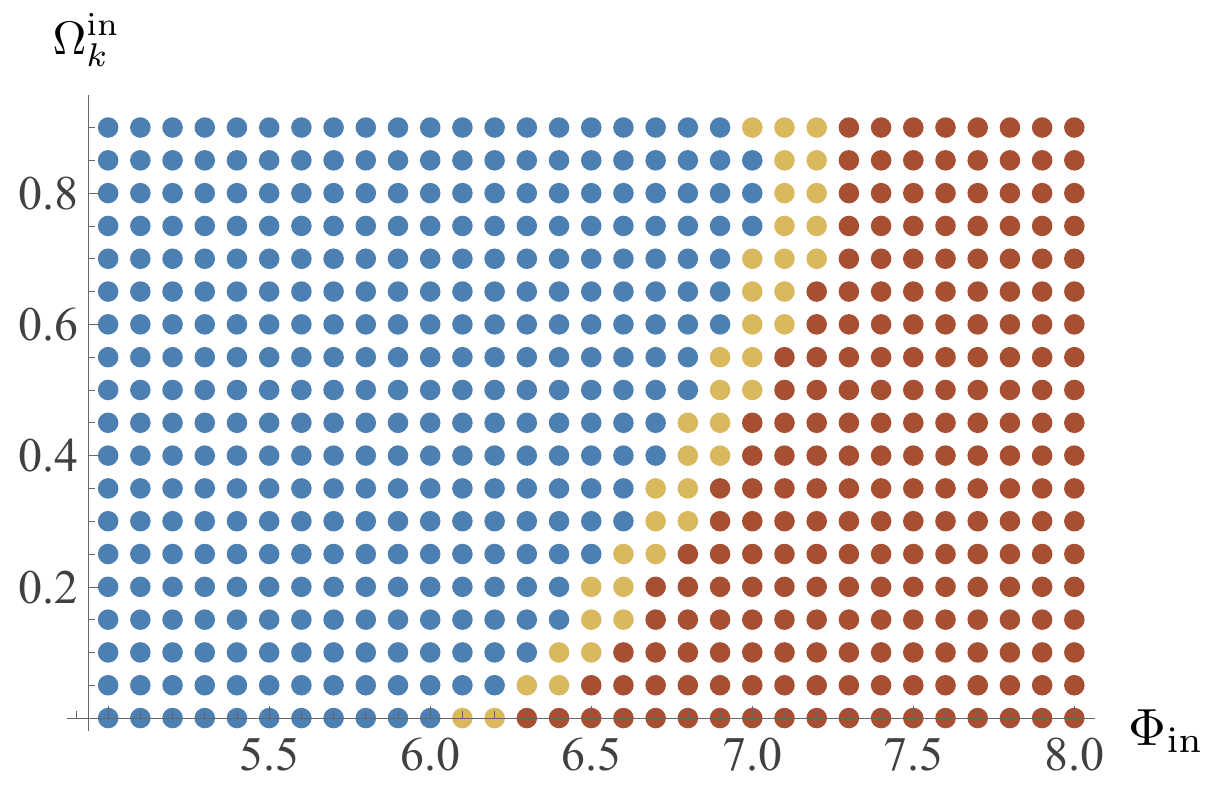}
\caption{Scan of initial conditions for $\Phi_\i$ and $\Omega_k^{\i}$ with $\Omega^{\i}_{\rm loop}=4.5 \times 10^{-3}$, $\Omega_{\rm rad}^{\i}= 10^{-3}$ and equal repartition among loop species. Blue dots correspond to stabilisation in the absence of radiation, yellow dots to stabilisation thanks to radiation, while red dots to overshooting.}
\label{fig:X_Phi_Rad_Low}
\end{figure}

In Fig. \ref{fig:X_Phi_Rad_Low} we performed the same scan as in Fig. \ref{fig:X_Phi Scan No Fluid low loops} with the same initial conditions in loops, but adding a $0.1\%$ initial energy density in radiation. 

\begin{figure}[h]
    \centering
    \includegraphics[width=\textwidth,
    height = 0.3 \textwidth]{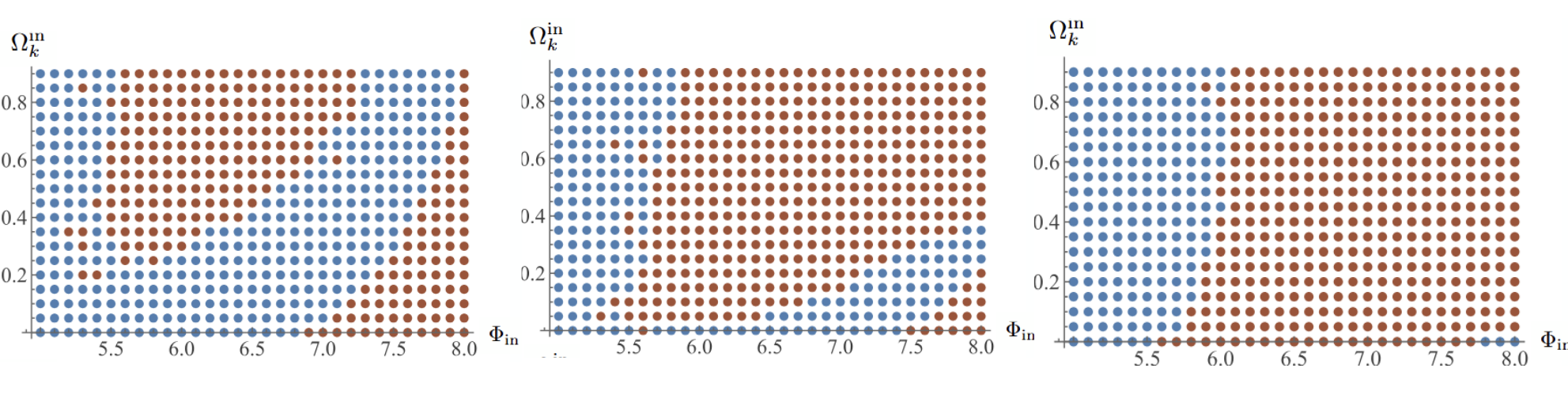}
\caption{Scan of initial conditions for $\Phi_\i$ and $\Omega_k^{\i}$ for various values of $\Omega^{\i}_{\rm rad}$. In each case $\Omega^{\i}_{\rm loop} = 4.5 \times 10^{-3}$ and $\Omega_j^{\i} = \Omega^{\i}_{\rm loop}/3$. Blue dots correspond to stabilisation, while red dots to overshooting. Left to right: $\Omega_{\rm rad}^{\i} =0.005\,, 0.01\,, 0.02$.}
\label{fig:X Phi various Omegarad}
\end{figure}

Increasing the amount of radiation while keeping the fraction of loops fixed, we find a similar phenomenon as that displayed in Fig. \ref{fig:X Phi various Omegaloop}. In Fig. \ref{fig:X Phi various Omegarad}, we can see that, surprisingly, increasing the fraction of radiation tends to destabilise the system. The reason for this behaviour is a triple interference between the energy densities of NS5-strings, D3-strings and radiation. Increasing the fraction of energy density in radiation, the time spent by the system closer to the fluid-dominated fixed point increases. The system then moves towards its final attractor, i.e. $\mathcal{T}_{1}^{(5)}$. In doing so, the field undergoes acceleration, also visible in Fig. \ref{fig:energy density with radiation} around $N = 9$. If this acceleration happens too close to the minimum, NS5-strings do not have enough time to absorb energy from the rolling field and brake it. This corresponds to the time when radiation becomes subdominant with respect to NS5-strings. The result is a system which is extremely sensitive to the initial energy densities in loops and radiation.

\begin{figure}[h]
    \centering
    \includegraphics[width=0.65\textwidth]{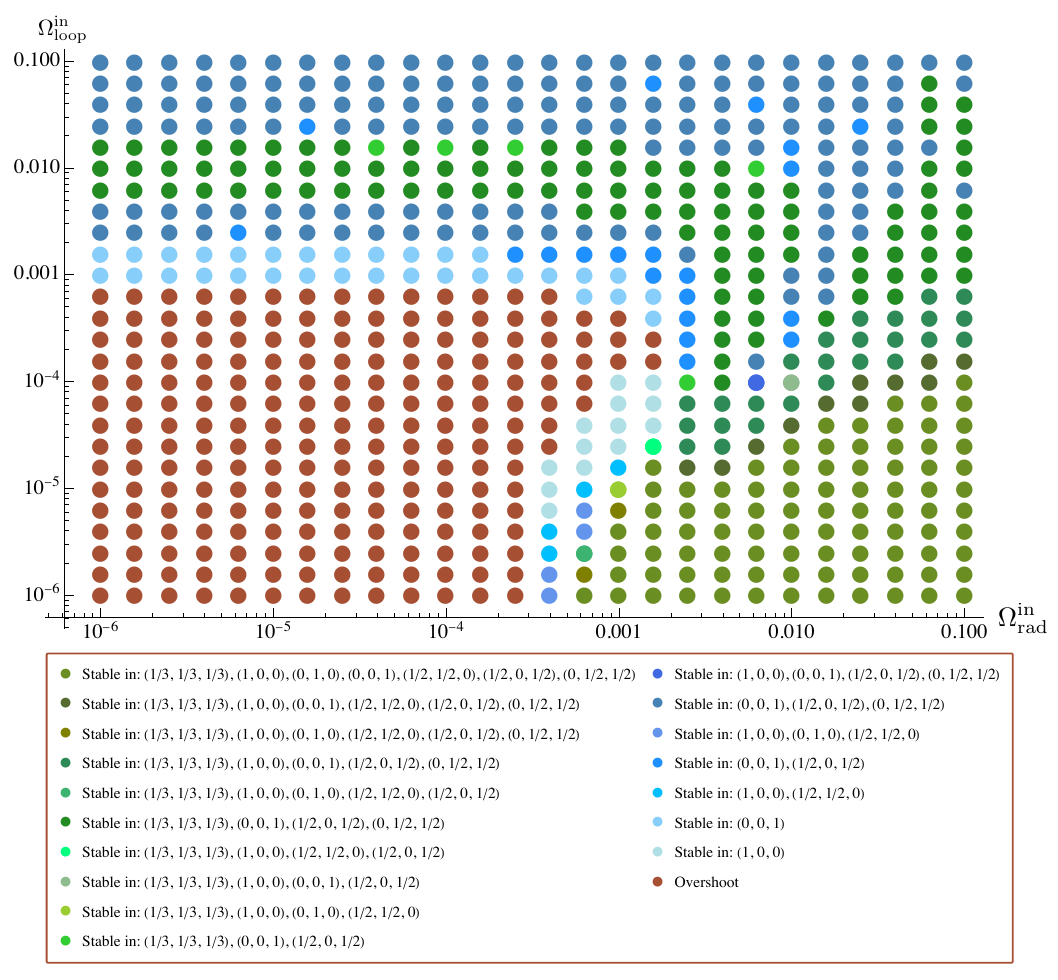}
\caption{Stabilisation map of the field with radiation and multiple loop species. The vertical axis is on a logarithmic scale and in each case we set $\Omega_k^{\i} = 0$. Different colours indicate different fractions of energy density in each loop species $\left(\Omega_F^{\i}, \Omega_3^{\i}, \Omega_5^{\i}\right)/\Omega_{\rm loop}^{\i}$.}
    \label{fig:Omega_loop vs Omega_rad}
\end{figure}

In Fig. \ref{fig:Omega_loop vs Omega_rad}, we show a scan over the initial fraction of energy density in radiation and loops fixing the initial conditions $\Phi_\i = 6 M_p$ and $\dot \Phi_\i = 0$. We can distinguish a few `asymptotic' regimes, that can be roughly identified with the corners of Fig. \ref{fig:Omega_loop vs Omega_rad}: 
\begin{itemize}
\item In the bottom-left corner ($\Omega_{\rm rad}^{\i} \lesssim 10^{-4}$, $\Omega_{\rm loop}^{\i}\lesssim 10^{-4}$) lies the low-loop, low-radiation regime. As expected, here the system behaves as in the absence of radiation, and the field always overshoots.
    
\item In the bottom-right corner ($\Omega_{\rm rad}^{\i} \gtrsim 10^{-2}$, $\Omega_{\rm loop}^{\i}\lesssim 10^{-4}$), we are in the low-loop, high-radiation regime. Here the field stabilises only due to the friction caused by the background radiation, and the loop distribution seems to play no role, as indicated by the fact that the field stabilises for all combinations of initial conditions.

\item In the top-left corner ($\Omega_{\rm rad}^{\i} \lesssim 10^{-3}$, $\Omega_{\rm loop}^{\i}\gtrsim 10^{-2}$) there is the high-loop, low-radiation limit, where the system essentially reduces to the case without radiation and a high loop fraction. In fact, the system here stabilises when we have a high fraction of NS5-strings. 

\item The top-right corner, as well as the central part, are a regime where the energy densities of the string loops and that of radiation are comparable in fraction. Here we have interference effects, similar to those in Fig. \ref{fig:X Phi various Omegarad}, due to the competition among radiation and strings which ultimately gives rise to an inefficient braking effect on the field.  
\end{itemize}

\section{Gravitational waves from multiple cosmic superstrings}
\label{sec:GW Emission}

Let us now include in our analysis the GW emission from the cosmic string loops. We will assume the standard form of the emitted power per string loop \cite{Vachaspati:1984gt, Vilenkin:2000jqa}:
\begin{equation}
\label{eq:emitted power}
P_{\rm GW}= \frac{d E_{\rm GW}}{dt} =  \Gamma G \mu^2\,,
\end{equation}
where $G = \left(8 \pi M_p^2\right)^{-1}$ is Newton's constant and $\Gamma$ a constant, dimensionless parameter whose value is typically extrapolated by numerical simulations.\footnote{We stress that, at the time of writing, no numerical simulation of the dynamics of cosmic strings with time-dependent tension has been performed, and so the values of $\Gamma$ in the literature come exclusively from simulations with constant tension.}

Let us comment that in the case of a time-dependent tension, \eqref{eq:emitted power}  may, in principle, be modified  by terms proportional to $\dot \mu$. The precise evaluation of the effect of these terms on the overall GW emission is an interesting question \textit{per se}, and will be explored in a future work. For now, we simply point out that in a properly defined \textit{adiabatic} regime of the variation of the tension, in which the period of oscillation of a loop, $T$, is much smaller than the timescale of the variation of the tension, the GW emission should not be much affected. In fact, this happens when:
\begin{equation}
\label{eq:adiabaticity}
    \left|\frac{\dot \mu}{\mu}\right| T \ll 1\,.
\end{equation}
From \eqref{eq:exponential tension}, $|\dot \mu/\mu| = \sqrt{6} \beta |\dot \Phi|$ in Planck units. From \eqref{eq:Friedman constraint}, $ |\dot \Phi| \leq \sqrt 6 H $. On the other hand, the period of oscillation of a loop is of order its physical length $T \simeq \ell$. Therefore we get:
\begin{equation}
\left|\frac{\dot \mu}{\mu}\right| T \lesssim 6 \beta H \ell \sim \O(1)\, H\ell\,.
\end{equation}
Therefore, given that we focus on the regime of small and isolated string loops characterised by $H\ell \ll 1$, the adiabaticity condition \eqref{eq:adiabaticity} is satisfied, justifying the use of the expression \eqref{eq:emitted power}.\footnote{The time-decrease of the tension of a cosmic superstring due to moduli dynamics is also bounded by UV constraints \cite{Brunelli:2026qkp}.}

The variation of the energy of one string loop, then, has two contributions: the variation of the tension, as in Sec. \ref{sec:NoEmission}, and the energy loss due to GW emission:
\begin{equation}
\frac{d E_{\rm loop}}{dt} = E_{\rm loop} \frac{1}{2 \mu} \frac{d \mu}{dt}- \Gamma G \mu^2\,.
\end{equation}
Recalling that $E_{\rm loop} = \ell\,\mu$ from (\ref{Eloop}), we can phrase this in terms of the variation of the string physical length $\ell$:
\begin{equation}
\label{eq:dl/dt GW}
\frac{d \ell}{dt} = - \ell \frac{1}{2\mu } \frac{d \mu}{dt}- \Gamma G \mu\,.
\end{equation}
Differentiating \eqref{eq:rho_loop} with respect to time we get:
\begin{equation}
    \dot \rho_{\rm loop} = -3H\rho_{\rm loop} + \frac{\dot \mu}{\mu} \rho_{\rm loop}+ \frac{\dot \ell}{\ell} \rho_{\rm loop}\,,
\end{equation}
and substituting in \eqref{eq:dl/dt GW}:
\begin{equation}
\label{eq:d rho/dt GW}
    \dot \rho_{\rm loop} = - 3H \rho_{\rm loop} + \frac{1}{2} \frac{\dot \mu}{\mu} \rho_{\rm loop}- \frac{\Gamma G \mu}{\ell} \rho_{\rm loop}\,.
\end{equation}
In the next subsections, we show how we can implement this directly in our dynamical system and study the evolution of the system and the overshoot problem in the presence of GW emission. 

\subsection{The dynamical system with GW emission}

Let us now translate this in the language of dynamical systems. Differentiating the definition \eqref{eq:Z} of the loop fractional energy density with respect to time, and using \eqref{eq:d rho/dt GW} and \eqref{eq:second Friedman}, we get:
\begin{equation}
\label{eq:Zs' GW}
    Z_j ' = \frac{3}{2} Z_j \left[X^2- Y^2 +\frac{W^2}{3} - \beta_j X - \frac{\delta_j}{3}\right],
\end{equation}
where we defined the dimensionless emission rate for the $j$-th loop species as:
\begin{equation}
\label{eq:delta}
    \delta_j = \frac{\Gamma G \mu_j}{H \ell_j}\,.
\end{equation}
Note that $\delta_j$ is not a constant, but rather a new variable (actually, a set of $N_s$ variables) of our dynamical system, whose dynamics crucially depends on the evolution of $\mu_j$, $\ell_j$ and $H$. When $\delta_j \sim \O(1)$, the (dimensionful) decay rate of the string loops into GWs is of order $H$, indicating that the decay becomes highly efficient and dominates the loop dynamics from that point onward. Another equation that gets modified is that of the radiation fluid.\footnote{Note that if, instead of a radiation background, we had a fluid with generic equation of state parameter $\omega$, we should include a new variable, $W_{\rm GW}$, parametrising the fraction of energy density in GWs. Since we will consider a radiation background, we will stick to a single fluid variable, $W$.} As the string loops decay into GWs, the energy density of the radiation fluid increases by:
\begin{equation}
\dot\rho_{\rm GW} \supset \sum_{j} \frac{\Gamma G \mu_j}{\ell_j}\, \rho_j\,.
\end{equation}
Including this effect in the equation for $W$, we get:
\begin{equation}
\label{eq:W' GW}
    W' = \frac{3}{2}W \left[-\frac{1}{3}+ \frac{1}{3 W^2}\sum_{j} \delta_j Z_j^2 +X^2 - Y^2+ \frac{W^2}{3}\right].
\end{equation}
Finally, we shall include in the dynamical system the equation for $\delta_j$. From \eqref{eq:delta}  we get:
\begin{equation}
    \dot \delta_j = \frac{\dot \mu_j}{\mu_j} \delta_j-\frac{\dot H}{H} \delta_j- \frac{\dot \ell_j}{\ell_j} \delta_j\,.
\end{equation}
Substituting in \eqref{eq:exponential tension}, \eqref{eq:second Friedman} and \eqref{eq:dl/dt GW}, and translating in number of efoldings, we end up with:
\begin{equation}
\label{eq:delta'}
\delta_j' = \delta_j \left[(1-9\beta_j X) \delta_j + \frac{3}{2} \left(1+ X^2 - Y^2+\frac{W^2}{3} \right)\right].
\end{equation}
The equations for $X'$ and $Y'$ are instead unchanged with respect to \eqref{eq:X'}-\eqref{eq:Y'}.

We shall now investigate the proper initial conditions for $\delta_j$. While independent as a dynamical variable, the initial conditions for the emission rate are constrained by the requirement that the string loops are \textit{subhorizon}. This condition guarantees that treating the cosmic strings as loops, rather than as a network, is well justified. Mathematically, it amounts to requiring a diluted fluid of loops with $H  \ell_j\ll 1$, exactly the combination that enters the definition of $\delta_j$. The second parameter entering the initial condition for $\delta_j$ is $G \mu_j$. In our setup this is not free, but set by the initial condition for the rolling modulus:
\begin{equation}
\label{eq:initial tension}
    \mu_j^{\i} = \tilde{\mu}_j \,  e^{-\sqrt{6} \,  \beta_j \Phi_\i}\,,
\end{equation}
where $\tilde{\mu}_j$ is the tension of the brane wrapping the cycle (or the fundamental string tension), with the dependence on the volume factored out. In general, $\tilde{\mu}_j$ depends on the string coupling $g_s$, which can accommodate a limited amount of tuning (see \cite{Chauhan:2025rdj, Chauhan:2026gid,Chauhan:2026dss} for recent scans of type IIB flux vacua). Therefore, for a selected value of $\Phi_\i$, in general $\delta_j^{\i}$ can only be made \textit{larger} by reducing $H_\i \ell_j^{\i}$, as this can be arbitrarily small, but is bounded from above by $1$.\footnote{Note that, in realistic scenarios, not all loops will have the same length, but there will be a distribution $P(\ell, t)$ depending on the details of the production mechanism. An example of such a distribution is given in Sec. \ref{sec:GW_Spectrum}.} For F-strings, D3-strings and NS5-strings, the initial values of $\delta_j$ in terms of the initial field value $\Phi_{\rm in}$ and string coupling $g_s$ are:
\begin{eqnarray}
\delta_F^{\i}&=& \frac{\Gamma \sqrt{g_s}}{16 \pi \,  H_\i \ell_F^{\i}}\, e^{-\sqrt{\frac{3}{2}}\, \Phi_\i}\,,\\
\delta_3^{\i}&=& \frac{\Gamma} {16 \pi  \,H_\i \ell_3^{\i}} \,e^{-\sqrt{\frac{2}{3}}\, \Phi_\i}\,,\\
\delta_5^{\i}&=& \frac{\Gamma }{16 \pi \, \sqrt{g_s} \,  H_\i \ell_5^{\i}} \,e^{-\frac{1}{\sqrt 6}\, \Phi_\i}\,.
\end{eqnarray}
For values of $\Phi_\i$ sufficiently small (less than around 14 in Planck units for $g_s \sim H_0\ell_5^{\i}\sim 0.1$) and $\Gamma = 50$, $\delta_5^{\i}$ is always of order $1$. 
This implies that NS5-strings will evaporate very quickly, and will not contribute much to the dynamics of the system, acting only as an initial seed of (gravitational) radiation. This dramatically changes the overshoot picture, as NS5-strings were those that drained energy from the field most efficiently, contributing most to the Hubble friction, since they redshifted the slowest. Without NS5-strings one may think that the field cannot stabilise at its minimum. However, in the next section we will see that the GWs produced by the decay of NS5- and D3-strings behave as background radiation which acts as a friction that prevents overshooting, while reaching a very large percentage of the energy density of the Universe.

\subsection{The overshoot problem with GW emission}
\label{sec:Overshoot GW}

As stated in the previous section, the quick evaporation of NS5-strings drastically changes the overshoot picture. In fact, ignoring dissipation into GWs, the string loops continuously exchange energy with the rolling modulus through the coupling in their tension. This interaction efficiently drains kinetic energy from the field, allowing slowly redshifting loop species, in particular NS5-strings, to prevent the modulus from overshooting. Once GW emission is included, the evolution of the system is determined by the competition between two distinct processes. On the one hand, the modulus transfers energy to the loop fluids through the coupling proportional to $\beta_i$. On the other hand, the string loops continuously lose energy through GW emission, increasing the radiation energy density. The latter redshifts more slowly than the kinetic energy of the modulus, so that the produced GWs provide an additional source of Hubble friction. 

An important aspect of the dynamics is that different cosmic string species do not decay at the same rate. For identical values of $\Gamma$, the emission parameter $\delta_j$ depends on both the string tension and its length, as seen in \eqref{eq:delta}, leading to different evaporation times for the various species. In particular,  D3- and NS5-string loops are heavier than F-strings due to their lower $\beta$'s, so that they lose energy more efficiently. Consequently, while heavier species rapidly transfer most of their energy into GWs, a substantial fraction of the F-string population survives for much longer.

This hierarchy has important consequences for the evolution close to the minimum of the potential. As discussed in the previous section, cosmic string fluids with larger values of $\beta_j$ tend to return part of their energy to the rolling modulus once the field approaches the barrier separating the minimum from the runaway region, thus causing overshoot.  Since F-strings remain the dominant surviving species at late times and possess the largest coupling among the string species considered here, they provide the largest contribution to this energy transfer. If their abundance becomes sufficiently large, this late-time injection of kinetic energy can compensate for the additional Hubble friction generated by the GWs, causing the field to cross the barrier and overshoot. We therefore expect the initial abundance of F-strings to play a crucial role in determining the final fate of the system.

The initial abundance of F-strings is constrained in two independent ways. First, their fractional energy density cannot become too large compared to that of the D3- and NS5-string populations. Otherwise, after the heavier species have efficiently decayed into GWs, the surviving F-string component dominates the loop sector and becomes the main source of energy exchange with the rolling modulus. Second, $\Omega_F^{\i}$ in itself should not exceed a certain threshold. In fact, even if the relative fraction of F-strings is kept fixed, increasing $\Omega_F^{\i}$ enhances the late-time transfer of energy back to the scalar field, eventually overcoming the friction provided by the generated GW background.

\begin{figure}[h]
    \centering
    \begin{subfigure}{0.49 \textwidth}
        \centering \includegraphics[width=\textwidth]{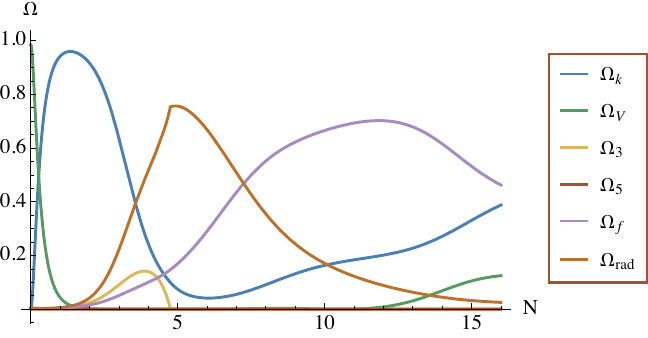} 
    \end{subfigure}
    \begin{subfigure}{0.49\textwidth}
        \centering
\includegraphics[width=\textwidth]{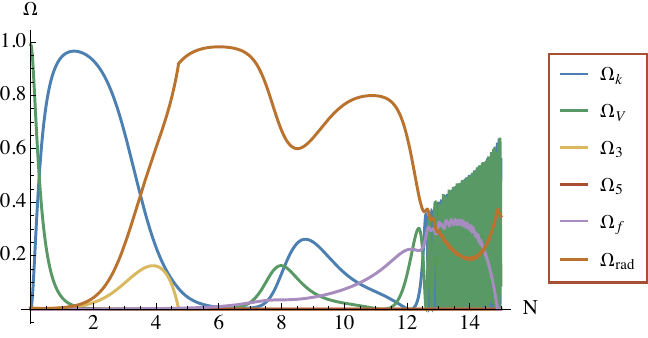}
 \end{subfigure}
\caption{Evolution of the energy densities for $\Gamma=50$, $H_\i\ell_F^{\i}=H_\i\ell_3^{\i}=H_\i\ell_5^{\i}=0.1$, and fixed D3- and NS5-string initial abundances. In the left panel all three species satisfy $\Omega_{F}^{\i}=\Omega_{3}^{\i}=\Omega_{5}^{\i}=0.005$ and the field overshoots. In the right panel only the F-string abundance is reduced to $\Omega_{F}^{\i}=10^{-4}$, leading instead to stabilisation.}
\label{fig:Frelative}
\end{figure}

To demonstrate these two effects separately, we consider two representative sets of initial conditions. In the first example, shown in Fig.~\ref{fig:Frelative}, all three loop species are initially assigned the same fractional energy density:
\begin{equation}
\Omega_{F}^{\i}=\Omega_{3}^{\i}=\Omega_{5}^{\i}=0.005\,,
\end{equation}
for which the field overshoots its minimum. We then reduce only the initial F-string abundance to:
\begin{equation}
\Omega_{F}^{\i}=10^{-4}\,,
\end{equation}
while keeping the D3- and NS5-string fractions unchanged. The resulting evolution no longer exhibits overshoot. This comparison shows that stabilisation is highly sensitive to the relative abundance of F-strings within the loop sector. Once the F-string component ceases to dominate the late-time dynamics, the energy extracted from the rolling modulus by the remaining loop species and the additional Hubble friction generated by GWs are sufficient to stabilise the field.

\begin{figure}[htbp]
    \centering
    \begin{subfigure}{0.49 \textwidth}
        \centering \includegraphics[width=\textwidth]{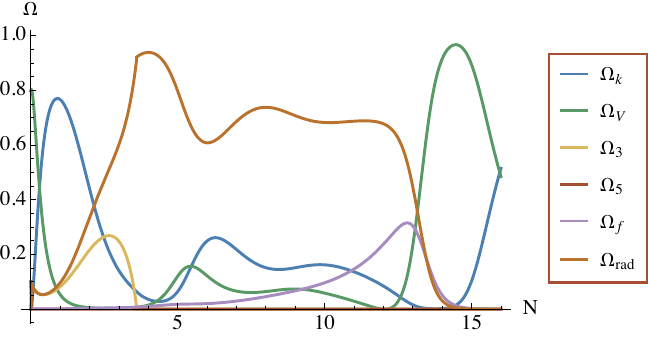} 
    \end{subfigure}
    \begin{subfigure}{0.49 \textwidth}
    \centering
        \includegraphics[width=\textwidth]{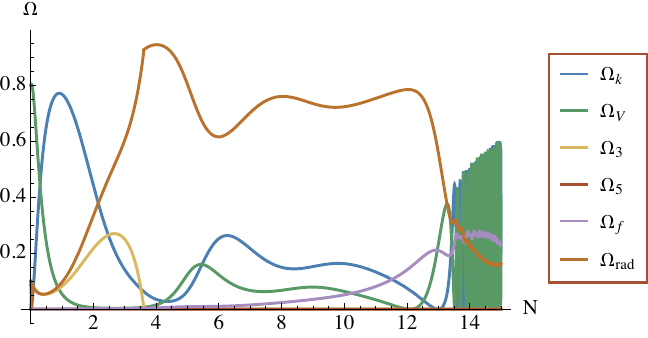}
    \end{subfigure}
    \caption{Evolution of the energy densities for $\Gamma=50$, $H_\i\ell_F^{\i}=H_\i\ell_3^{\i}=H_\i\ell_5^{\i}=0.1$, and fixed $\Omega_{3}^{\i}=\Omega_{5}^{\i}=0.1$. The left panel corresponds to $\Omega_{F}^{\i}=0.002$ and exhibits overshooting, while the right panel shows that reducing the initial F-string fraction to $\Omega_{F}^{\i}=0.001$ restores stabilisation.}
    \label{fig:Fabsolute}
\end{figure}

The second example illustrates that the absolute abundance of F-strings is equally important. Here we fix:
\begin{equation}
\Omega_{3}^{\i}=\Omega_{5}^{\i}=0.1\,,
\end{equation}
and compare two different initial F-string fractions. For:
\begin{equation}
\Omega_{F}^{\i}=0.002\,,
\end{equation}
the field overshoots, whereas reducing the F-string abundance to:
\begin{equation}
\Omega_{F}^{\i}=0.001\,,
\end{equation}
is sufficient to recover stabilisation, as shown in Fig.~\ref{fig:Fabsolute}. Since the relative abundance of the heavier species is identical as to the one in Fig. \ref{fig:Frelative}, the different outcome is entirely due to the larger amount of energy stored initially in the F-string population. This confirms that the late-time dynamics is controlled not only by the relative hierarchy among the loop species, but also by the absolute fraction of energy density of the long-lived F-strings.

\begin{figure}[htbp]
    \centering
    \includegraphics[width=0.65\textwidth]{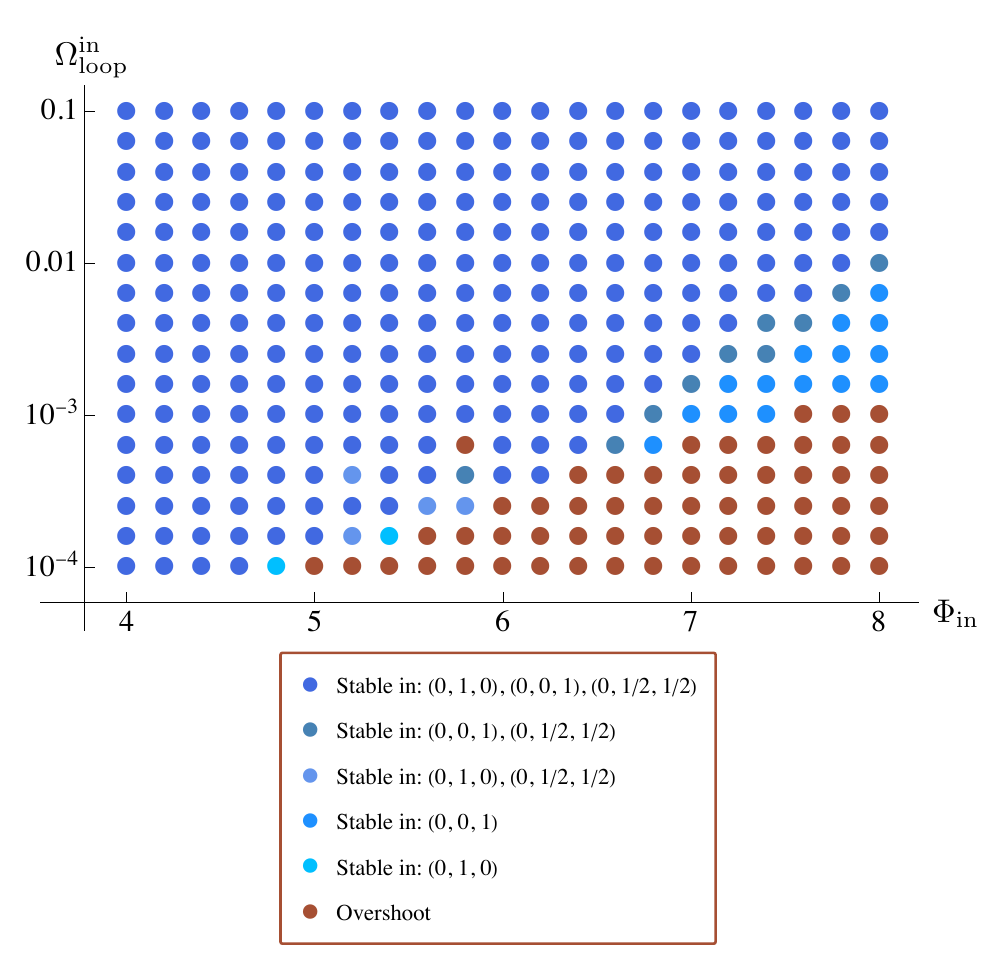}
    \caption{Stability maps in the $(\Phi_\i,\Omega_{\rm loop}^{\i})$ plane for $\Gamma=50$ and $H_\i\ell_F^{\i}=H_\i\ell_3^{\i}=H_\i\ell_5^{\i}=0.1$, considering different initial loop configurations. The vertical axis is on a logarithmic scale. Blue points correspond to stabilisation, while red points to overshooting.}
    \label{fig:stability_maps_Fstrings}
\end{figure}

To better understand the role played by the initial F-string abundance, we performed a scan over the initial conditions in the $(\Phi_\i,\Omega_{\rm loop}^{\i})$ plane. The resulting stability maps for different initial loop configurations are shown in Fig.~\ref{fig:stability_maps_Fstrings}. A striking feature immediately emerges from these scans. Whenever the initial F-string abundance is chosen to be comparable to that of the D3- and NS5-string populations, we do not find any region of parameter space where the modulus successfully stabilises. In all such configurations the field inevitably overshoots its minimum. This behaviour is a direct consequence of the long lifetime of the F-string component. After the heavier loop species have efficiently decayed into GWs, the remaining F-strings continue to exchange energy with the rolling modulus, injecting kinetic energy back into the scalar field and eventually driving it beyond the potential barrier. Successful stabilisation is recovered only when the initial abundance of F-strings is sufficiently suppressed with respect to the other loop species.

Another important aspect that emerges from this analysis is that these evolutions typically feature a period of \textit{gravitational wave domination}, as seen in Figs. \ref{fig:Frelative} and \ref{fig:Fabsolute}. Such large backgrounds of GWs at the time of their emission are  a general prediction of this model. In the following section, we investigate what observational consequences these have on the stochastic GW background observed today. 

One could also be worried that, if such a large energy density is stored in GWs at any given time during the system evolution, the perturbative expansion of the metric that gave rise to the coupling with the GWs in the first place may be out of control. However, perturbation theory is safe due to the fact that the high energy of these GWs is accounted for by their high frequency, rather than their amplitude. A detailed calculation of the energy-momentum tensor of these GWs is given in App. \ref{app:GW tensor}.

\subsection{Gravitational wave spectrum}
\label{sec:GW_Spectrum}

Having described the evolution of multiple species of cosmic superstrings and their GW backreaction, we now compute the present-day GW spectrum sourced by their decay.

For this purpose, we can solve the dynamical system formed by the equations \eqref{eq:X'}, \eqref{eq:Y'}, \eqref{eq:Zs' GW}, \eqref{eq:W' GW} and \eqref{eq:delta'}, together with \eqref{eq:second Friedman} to get the time-evolution of $H$. 

\begin{figure}[h!]
\centering
\includegraphics[width=0.65\textwidth]{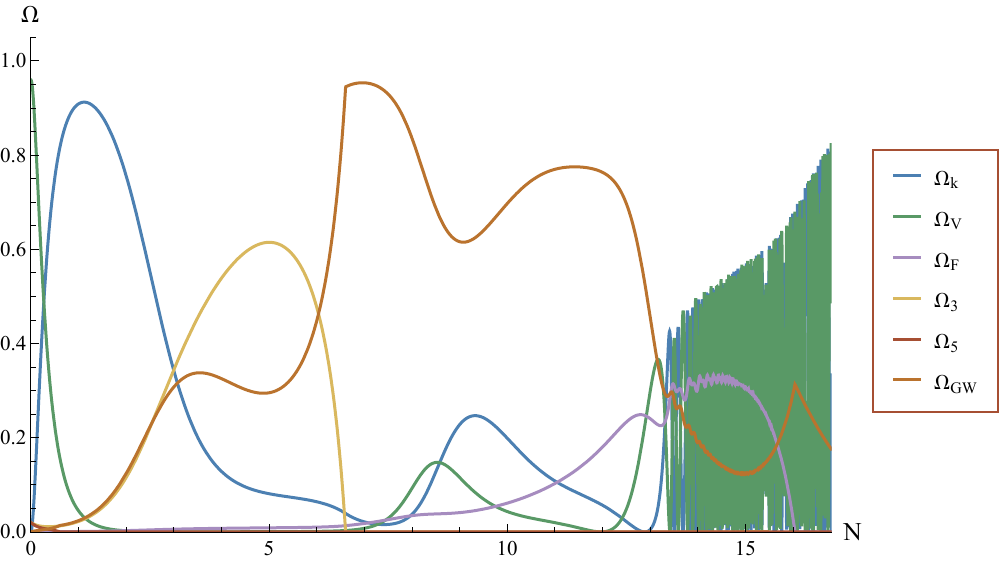}
\caption{Evolution of the energy densities with initial conditions $\Phi_\i = 6M_p$, $\dot{\Phi}_\i = 0$, $\Omega_{5}^\i= \Omega_{3}^\i= 2 \times10^{-2}$ and $\Omega_F^\i = 10^{-3}$. Initial sizes for all strings are $H_{\rm in}\ell_{F }^\i = 0.3$, $H_{\rm in}\ell_{3}^\i = 0.3$ and $H_{\rm in}\ell_{5}^\i = 0.3$ with  $g_s = 0.3$.}
\label{fig:DensityPlot_4}
\end{figure}

The main difference with respect to \cite{Conlon:2025mqt} is that, in the solutions considered here, the dominant GW emission for D3- and NS5-strings occurs before the onset of modulus domination. In order to plot the spectrum, we take a particular case of the dynamics with reasonable initial conditions, i.e. subhorizon loops ($H\ell \ll 1$) and  small initial fraction of energy density in loops, as shown in Fig. \ref{fig:DensityPlot_4}. Note, in particular, that the energy density in F-strings before their decay is about $1/3$ of the total.

Under the assumption of non-interacting loops, we will compute the GW spectrum for each string species separately by generalising (while also reviewing) the approach of \cite{Conlon:2025mqt}, allowing the cosmic strings to decay before matter domination, when the tension is still varying. The GW energy density fraction observed today is given by:
\begin{equation}
\label{eq:basicSpectrum}
    \Omega^0_{\rm GW}(f_{\rm obs}) =\frac{f_{\rm obs}}{\rho_{c,0}} \int_{t_{\rm in}}^{t_f} \frac{d{\dot{\rho}_{\rm em}}}{df_{\rm em}} \left(\frac{a(t)}{a(t_0)}\right)^3 \, dt\,,
\end{equation}
where, $\rho_{c,0} = 3 M_p^2 H^2_0$ is the critical energy density today and $f_{\rm obs}$ is the observed frequency. Essentially, we are integrating the energy density emitted per unit time per unit frequency, taking into account the redshift  until today. Here,  $t_{\rm in}$ is the initial time of emission of GWs from the cosmic strings, and $t_f$ indicates the final time.

The energy emitted per unit time per unit of frequency is given by:
\begin{equation}
\label{eq:energyPerTimePerFreq}
\frac{d{\dot{\rho}_{\rm em}}}{df_{\rm em}} = \sum_k \frac{dE^{(k)}}{dt}\frac{dn_{\rm loop}(\ell,t)}{df_{\rm em}}\,,
\end{equation}
where $dE^{(k)}/dt$ indicates the power emitted in the $k$-th harmonic mode, and $n_{\rm loop}(\ell, t)$ is the number density distributions of loops as a function of time and length. Let us assume the lengths of the loops are distributed according to a given distribution $P(\ell, t)$. Then we can write:
\begin{equation}
\frac{d n_{\rm loop}(\ell, t)}{df_{\rm em}} = \frac{dn_{\rm loop}(\ell,t)}{d\ell} \left| \frac{d \ell}{df_{\rm em}} \right|= n_{\rm loop}(t) P(\ell, t) \left|\frac{d \ell}{df_{\rm em}}\right|\,,
\end{equation}
where the $n_{\rm loop}$ factor accounts for the redshift of the loop population and the absolute value is taken to ensure the positivity of the distribution.  The length of a loop is related to the frequency of the $k$-th mode as: 
\begin{equation}
\label{eq:freqLength}
    f_{\rm em} = \frac{2k}{\ell(t_{\rm em})}\,,
\end{equation}
where the length evolves with time as illustrated in \eqref{eq:dl/dt GW}. For each mode, we have that $dE^{(k)}/dt = \Gamma^{(k)} G \mu^2$, where $\Gamma^{(k)}$ is a dimensionless coupling, giving:
\begin{equation}
\frac{d{\dot{\rho}_{\rm em}}}{df_{\rm em}} = \sum_{k}\Gamma^{(k)}G\mu^2(t)\,  n_{\rm loop}(t_{\rm em})\, P(\ell, t)\, \left|\frac{d \ell}{df_{\rm em}}\right|\,.
\label{eq:FinalEnergyEmitted}
\end{equation}
Converting \eqref{eq:FinalEnergyEmitted} and \eqref{eq:basicSpectrum} into number of e-foldings $N$, with $dt = dN/H(N)$, we can write:
\begin{eqnarray}
\Omega^0_{\rm GW}(f_{\rm obs}) &=&
\frac{f_{\rm obs}}{\rho_{c,0}}
\sum_k \int_{N_{\rm in}}^{N_f} \frac{dN}{H(N)}\,
\Gamma^{(k)} G\mu^2\,
n_{\rm loop}(N_{\rm em}) \nonumber \\
&\times&
\, P(\ell, N) \left|\frac{\ell}{f_{\rm em}}\right|
\left(\frac{a(N)}{a_0}\right)^3 .
\label{eq:ObsSpecInt}
\end{eqnarray}
To go any further, we have to specify a length distribution. First, in a similar spirit as \cite{Conlon:2025mqt}, we suppose all the loops have the same length, so that:
\begin{equation}
\label{eq:same length}
    P(\ell, N)  = \delta\!\left(\ell-\bar \ell(N)\right).
\end{equation}
Going back to the frequency regime, we get: 
\begin{equation}
\label{eq:same length frequency}
    P(\ell(f_{\rm em}), N) = \left|\frac{f_{\rm em}}{\ell}\right| \delta\!\left(f_{\rm em}- \frac{2k}{\ell(N)}\right)
\end{equation}
Inserting \eqref{eq:same length} in \eqref{eq:ObsSpecInt} and using the properties of the delta function, we get:
\begin{eqnarray}
\Omega^0_{\rm GW}(f_{\rm obs}) 
&=&\frac{1}{\rho_{c,0}} 
\sum_k\int_{N_{\rm in}}^{N_f}\frac{dN}{H(N)} \Gamma^{(k)}G\mu^2 n_{\rm loop}(N) \nonumber \\
&\times& \frac{\delta(N - N_{\rm em})}{|\ell^\prime/\ell-1|}\left(\frac{a(N)}{a_0}\right)^4\,,
\label{eq: ObsSpecInt2}
\end{eqnarray}
where the denominator is the Jacobian introduced by simplifying the delta function, $N_{\rm em}$ now refers to the e-folding/time of emission, and we have used $a = e^{N}$. We have also used the fact that the observed frequency is related to the emission time by redshifting the emitted frequency as:
\begin{equation}
f_{\rm obs}= \frac{a(N_{\rm em})}{a_0}\frac{2k}{\ell(N_{\rm em})}\,.
\label{eq:ObsFreq}
\end{equation}
We can perform the integral because of the delta function, obtaining:
\begin{equation}
\Omega^0_{\rm GW}(f_{\rm obs}) = \sum_{k}\Omega^{\rm em,(k)}_{\rm GW}(N_{\rm em})\frac{H_{\rm in}/H(N_{\rm em})}{|\ell^\prime/\ell-1|}\, e^{(N_{\rm em}-N_{\rm in})}\,.
     \label{eq:ObserSpec}
\end{equation}
In \eqref{eq:ObserSpec}, we introduced $\Omega^{\rm em,(k)}_{\rm GW}(N_{\rm em})$ defined as:
\begin{eqnarray}
\Omega^{\rm em,(k)}_{\rm GW}(N_{\rm em}) 
&\equiv&  \frac{\Gamma^{(k)}G\mu(N_{\rm em})}{H_{\rm in} \ell _{\rm in}}\bigg(\frac{\mu(N_{\rm em})}{\mu_{\rm in}}\bigg) \nonumber \\
&\times& \Omega_{\rm loop}^{\rm in}\bigg(\frac{H_{\rm in}}{H_0}\bigg)^2 \bigg(\frac{a_{\rm in}}{a_0}\bigg)^4\,.
\label{eq:emitOmega}
\end{eqnarray}
When the field $\Phi$ reaches its minimum, it starts oscillating, giving rise to a period of early matter domination. This era ends when the modulus decays and reheats the universe, i.e. when $H \simeq \Gamma_\Phi$. It is therefore necessary to estimate the decay rate of the volume modulus. Moduli typically couple gravitationally, and their decay time tends to be very long due to Planck-suppressed couplings. Nevertheless, whenever the Standard Model is not sequestered from the sources of supersymmetry breaking and the volume mode is heavy enough to avoid the CMP, the volume mode features an enhanced decay width into Higgses of the form \cite{Cicoli:2022uqb}:
\begin{equation}
\label{eq:SVLLVS}
    \Gamma_\Phi \simeq (c\, \mathcal{V})^2 \frac{m_\phi^3}{M_p^2}\,,
\end{equation}
where $c$ is a loop factor and the mass of the volume modulus is $m_\phi \simeq M_p/\mathcal{V}^{3/2}$. We then multiply and divide \eqref{eq:emitOmega} by the Hubble rate at the time of the onset of matter domination, $H_\Phi^2$, and by  $\Omega_{\rm rad}^0 = (H_{\rm rh}/H_{0})^2(a_{\rm rh}/a_0)^4$, where the subscript `rh' indicates reheating time: 
\begin{eqnarray}
\Omega^{\rm em,(k)}_{\rm GW}(N_{\rm em})
&=&
\frac{\Gamma^{(k)}G\mu(N_{\rm em})}{H_{\rm in} \ell_{\rm in}}
\left(\frac{\mu(N_{\rm em})}{\mu_{\rm in}}\right)
\Omega_{\rm loops}^{\rm in} \nonumber \\
&\times&
\Omega_{\rm rad}^{0}
\left(\frac{H_{\rm in}^2\, H_{\rm rh}^{2/3}}{H_\Phi^{8/3}}
\right) e^{4(N_{\rm in}-N_\Phi)}\,,
\label{eq:emitOmegaFinal}
\end{eqnarray}
where we used the fact that, during matter domination, $a \sim t^{2/3}$. 
The factor $e^{4(N_{\rm in} - N_\Phi)}$ measures the redshift of the emitted GW radiation from the initial time to the onset of the modulus domination; on the other hand, $(H_{\rm rh}/H_{\Phi})^{2/3}$ captures the dilution of the gravitational radiation during the modulus domination period. 

\begin{figure}[h!]
\centering
\includegraphics[width=0.65\textwidth]{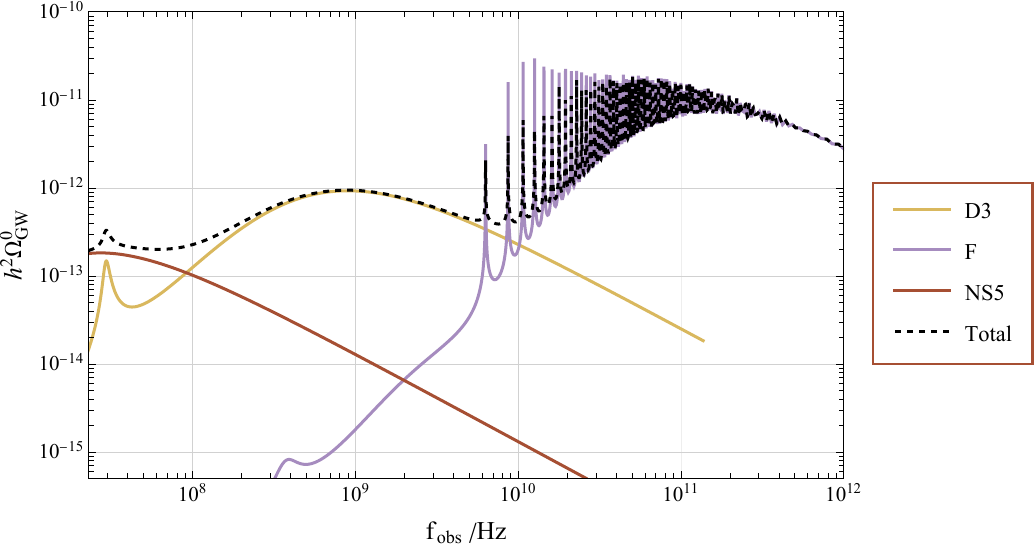}
\caption{GW spectrum from the three string loop species for the evolution in Fig. \ref{fig:DensityPlot_4}. The initial conditions are $\Phi_{\rm in} = 6 M_p$, $\dot{\Phi}_{\rm in} = 0$, $\Omega_{5}= \Omega_{3}= 2 \times10^{-2}$ and $\Omega_F = 10^{-3}$. Sizes for all strings are $H_{\rm in}\ell_{\rm F, \, in} = 0.3$, $H_{\rm in}\ell_{\rm 3, \, in} = 0.3$ and $H_{\rm in}\ell_{\rm 5,\, in} = 0.3$. We also take $g_s = 0.3$ and $c=1/(4\pi)$.}
    \label{fig:spectrum_4}
\end{figure}

Fig. \ref{fig:spectrum_4} shows the GW spectra produced by the three loop species for the benchmark evolution of Fig. \ref{fig:DensityPlot_4}, as observed today. The relative positions and amplitudes of the peaks are controlled mainly by the emission time of each species. Radiation emitted earlier is redshifted for a longer period before reheating, leading to a lower observed frequency and to a smaller amplitude. The NS5-string population decays first, and therefore gives the most redshifted contribution. The D3- and F-string spectra exhibit an additional bump in the spectrum associated with the early kination epoch, during which the background energy density redshifts faster than radiation, and the GW fraction is correspondingly enhanced. Finally, we do not take into account higher harmonics above $k=1$ as they do not change the peak behaviour of the spectrum but only flatten and extend the tail towards higher frequencies.

The sharp oscillatory features in the F-string spectrum arise close to the onset of modulus oscillations. In this regime the time-dependent tension induces rapid variations in the loop length, and the Jacobian factor $|1 - \ell^\prime/ \ell|$ in the denominator of \eqref{eq:ObserSpec} becomes small. These narrow peaks are due to the assumption of a monochromatic loop population. 

To test this assumption, we can modify our loop population \eqref{eq:same length} to be a finite initial length distribution. We shall remark that the precise length distribution is highly dependent on the loop production mechanism, which we ignore here. To model a physically-relevant smeared distribution, we choose a log-normal distribution in length: 
\begin{equation}
\label{eq: l_dist}
    P({\ell}, N) = \frac{1}{\sqrt{2\pi}\sigma \ell}\, e^{- \frac{1}{2\sigma^2}\ln^2( \ell / \bar\ell(N))}\,,
\end{equation}
where $\sigma^2$ is the variance of the distribution, and $\ln{(\bar\ell(N))}$ is the mean logarithmic length at a given emission time. Inserting this distribution in \eqref{eq:ObsSpecInt}, the spectrum becomes:
\begin{eqnarray}
\Omega^0_{\rm GW}(f_{\rm obs}) 
&=&\frac{1}{\rho_{c,0}} 
\sum_k\int_{N_{\rm in}}^{N_f}\frac{dN}{H(N)} \Gamma^{(k)}G\mu^2  \bigg(\frac{a(N)}{a_0}\bigg)^4 \nonumber \\
&\times& \frac{n_{\rm loop}(N)}{\sqrt{2\pi}\sigma}\, e^{-\frac{1}{\sigma^2}\left[\ln(f_{\rm obs}) - \ln(\bar f_{k}(N))\right]^2}\,, 
\label{eq: gen_ObsSpecInt_dist_final}
\end{eqnarray}
where we used  $ f_{\rm obs} = (a(N)/a_0) (2k/\ell)$ and we have defined:
\begin{equation}
\label{eq: f_0,k}
    \bar f_{k}(N) \equiv \frac{a(N)}{a_0} \frac{2k}{\bar \ell(N)}\,.
\end{equation}
From \eqref{eq: gen_ObsSpecInt_dist_final}, it becomes clear that a distribution of lengths becomes a smearing of frequency for our particular choice of distribution. Comparing \eqref{eq: gen_ObsSpecInt_dist_final} with \eqref{eq: ObsSpecInt2}, we can see that the term $\delta(N - N_{\rm em})/|\ell^\prime/\ell-1|$, which causes sharp caustic-like features, is replaced by a smooth Gaussian exponential function.

\begin{figure}[h!]
\centering
\includegraphics[width=0.65\textwidth]{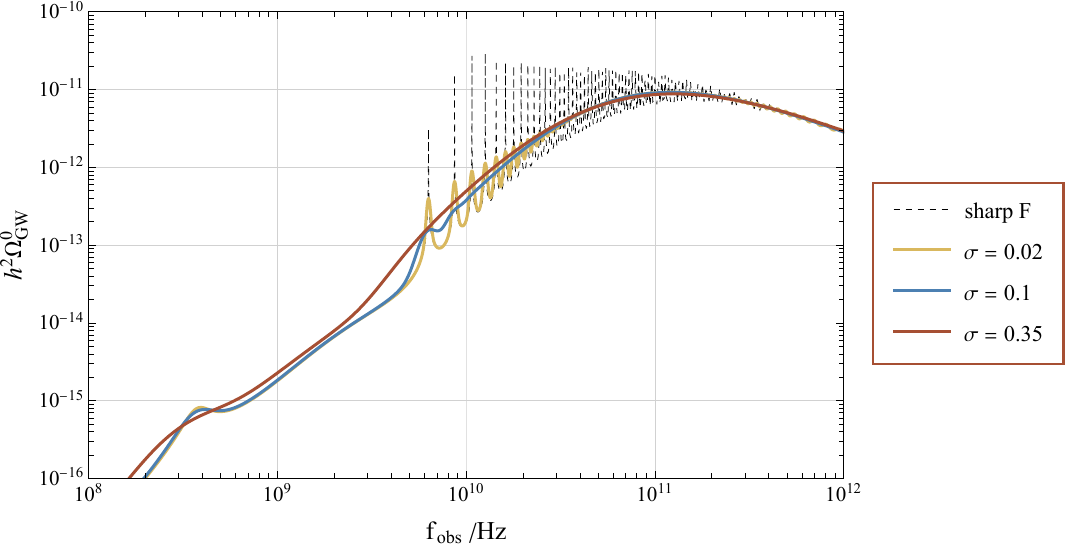}
\caption{GW spectrum sourced by the F-string loop population for the evolution of Fig.~\ref{fig:DensityPlot_4}. The dashed curve shows the monochromatic result, while the solid curves correspond to log-normal instantaneous length distributions with $\sigma=0.02$, $0.10$, and $0.35$. Increasing the width smooths the narrow spikes, while leaving the broad spectral maximum approximately unchanged.}
    \label{fig:smeared_spectrum}
\end{figure}

In Fig. \ref{fig:smeared_spectrum}, we can see a fair amount of smearing even for the smallest value $\sigma = 0.02$, and the spectrum becomes completely smooth for $\sigma = 0.35$. We see that the overall peak behaviour remains the same. We restrict to the fundamental mode, $k=1$. We do not recompute the coupled background evolution for a family of loop lengths. Instead, the central length $\ell_0(N)$ is taken from the single-length numerical solution, while a fixed logarithmic width is imposed phenomenologically at each emission time. A fully self-consistent treatment would evolve an initial length distribution, account for the different decay times of the subpopulations, and include their backreaction on the cosmological background (which we leave for future work). The present calculation should therefore be interpreted as a test of the sensitivity of the narrow spectral features to the monochromatic-loop approximation.

\section{Conclusions}
\label{Concl}

In this work we have investigated the cosmological evolution of multiple species of cosmic superstrings with time-dependent tension, and their impact on the stabilisation of a rolling modulus. In particular, we have studied the effect of GW emission from string loops and the resulting backreaction on the dynamics of the modulus. 

We have extended the dynamical system describing the coupled evolution of the rolling modulus and string loop fluids to include the energy loss due to GW emission. The validity of using the standard expression for the GW power emitted by a loop, $P_{\rm GW}=\Gamma G\mu^2$, has been justified in the adiabatic regime, which is naturally realised for subhorizon loops. In this regime the variation timescale of the string tension is much longer than the oscillation period of the loops, and corrections proportional to $\dot\mu$ do not significantly affect the emission rate. The energy lost by the loops is transferred into a stochastic GW background, whose contribution to the cosmological evolution can be consistently included as an additional radiation component.

A crucial consequence of GW emission is that the picture of the overshoot problem is qualitatively modified with respect to the case without radiation. In the absence of GW backreaction, heavy cosmic string species, and in particular NS5-strings, efficiently drain kinetic energy from the rolling modulus through their coupling to the varying tension. Once GW emission is included, however, these species rapidly evaporate, converting their energy into radiation. The resulting GW background can dominate the energy density of the Universe and provides an additional source of Hubble friction, allowing the modulus to stabilise even when the original NS5-string friction mechanism is no longer active.

Note that we treated the initial abundances and characteristic loop sizes as input parameters. Our analysis shows, however, a clear qualitative picture. In particular, the final outcome of the evolution is found to depend sensitively on the abundance of F-strings. Unlike D3- and NS5-strings, F-strings survive for much longer times and continue to exchange energy with the modulus close to the minimum of the potential. If their initial abundance is too large, the energy transferred back to the scalar field can overcome the friction generated by the GW background and lead to overshooting. Conversely, a sufficiently suppressed F-string population allows the friction from the GW radiation to stabilise the modulus. This highlights that the relative abundance of different cosmic string species is a key ingredient in determining a successful solution of the overshoot problem.

We have also computed the stochastic GW spectrum generated by the decay of the different string species, finding an interesting multi-peaked signal at high frequencies. The different string species leave distinct signatures in frequency space, determined primarily by their decay times and the subsequent redshift history. Earlier decaying species (NS5- and D3-strings) produce more redshifted signals, while the spectra of longer-lived F-strings can retain features associated with the non-standard cosmological eras preceding reheating.

The monochromatic approximation for the loop population leads to sharp spectral features associated with the Jacobian relating emission time and observed frequency. We have shown that these structures are sensitive to the assumed length distribution and are smoothed once a finite width distribution of loop sizes is considered. In particular, a log-normal distribution removes the narrow caustic-like peaks while leaving the broad spectral features and peak locations approximately unchanged. Given that the loop production function remains largely unknown, we stress that only the broad spectral features should be regarded as robust predictions, unlike the sharp oscillatory structures. Moreover, predicting the final GW spectrum in full detail would require a self-consistent evolution of a realistic initial distribution of loop lengths. This important analysis is beyond the scope of the present work, and so we leave it for future work.

Finally, we have addressed the possible concern that a GW background carrying an order-one fraction of the total energy density could invalidate the perturbative description of GWs. We have shown that this does not occur in the regime considered here. The large energy density is associated with the high frequency of the emitted GWs rather than with a large metric perturbation, and the amplitude of the tensor perturbations remains suppressed by the ratio $H/\omega\simeq H\ell\ll1$. Therefore, the perturbative treatment of the GW background remains under control.

Our results demonstrate that GW emission from cosmic superstrings is not merely a secondary energy loss effect, but can qualitatively alter the post-inflationary dynamics of moduli rolling over steep potentials towards a late time minimum. The interplay between the dynamical evolution of multiple species of cosmic superstrings, their decay into gravitational radiation, and the resulting cosmological GW background provides a very promising way to connect generic features of string cosmology to observations, via a stochastic GW signal that might be detected in the future. Let us stress that detecting stochastic gravitational waves at high frequencies is extremely challenging since it requires to go beyond conventional interferometric techniques. However, several ideas have been put forward including photon-GW conversion, remodulation of existing microwave-cavity experiments originally designed for axion searches, and quantum sensing \cite{Aggarwal:2025noe}.

A major obstacle is the dilution of the GW signal due to the epoch of matter domination caused by the oscillations of the modulus around the minimum of its potential. In this paper we considered the case when the modulus decay into Higgses is enhanced due to a large scale of supersymmetry breaking \cite{Cicoli:2022uqb}. It would be interesting to find additional mechanisms which could enlarge the amplitude of the GW spectrum by reducing the modulus lifetime.

\subsection*{Acknowledgments}

We would like to thank Edward Hardy, Filippo Revello, Noelia S\'anchez Gonz\'alez and Gonzalo Villa for useful discussions. The work of LB, MC and FGP contributes to the COST Action COSMIC WISPers CA21106, supported by COST (European Cooperation in Science and Technology).

\begin{appendix}

\section{Validity of perturbation theory for a large GW background}
\label{app:GW tensor}

Given the results presented in Sec. \ref{sec:Overshoot GW}, one might be worried that such a large GW background (more than $90 \%$ of the total energy density) may cause a breakdown of perturbation theory. In this section we show that this needs not be the case, and, in particular, that our analysis is safe due to the high frequency of the GWs at play. 

Gravitational waves are typically expressed as perturbations over a given background metric $\bar g_{\mu \nu}$:
\begin{equation}
\label{eq:perturbation condition}
    g_{\mu \nu} = \bar g_{\mu \nu}+ h_{\mu \nu} \quad \text{with} \quad |h_{\mu \nu}| \ll 1\,.
\end{equation}
Therefore, perturbation theory breaks down in a regime where $|h_{\mu \nu}| \sim 1$. We shall now give an argument why this is never the case in our system. Typically, the propagation of GWs is studied in the transverse traceless (TT) gauge defined as:
\begin{equation}
\de^\mu h_{\mu \nu}^{(TT)} =0 \qquad\text{and} \qquad h_{\lambda}^{\lambda \, (TT)} =0\,,
\end{equation}
where the trace is computed by raising the index with the background metric $\bar g_{\mu \nu}$. In this gauge, the energy-momentum tensor of the GW fluid at high frequencies is given by the Isaacson effective energy-momentum tensor, which, in a generic background spacetime can be written as: \cite{Isaacson:1968hbi, Isaacson:1968zza}:
\begin{equation}
\label{eq:Isaacson tensor}
    t_{\mu \nu}^{\rm (GW)} = \frac{1}{32 \pi G} \left \langle \overline\nabla_\mu h^{(TT)\,  \alpha \beta} \overline \nabla_\nu h^{(TT)}_{\,  \alpha \beta} \right \rangle\,, 
\end{equation}
where $\overline\nabla_\mu$ is the covariant derivative with respect to the background metric $\bar g_{\mu \nu}$ and angular brackets indicate spatial average. The energy density of GWs is therefore proportional to the $(0,0)$ component of this tensor:
\begin{equation}
\label{eq:rho_GW EM tensor}
    \rho_{\rm GW} \simeq t_{00}^{\rm (GW)} = \frac{M_p^2}{4} \left \langle \overline \nabla_0 h^{(TT) \, \alpha \beta }\, \overline \nabla_0 h^{(TT)}_{\alpha \beta}\right \rangle\,.
\end{equation}
The covariant derivative of the metric perturbation can be expanded as (we shall henceforth drop the (TT) superscript for the sake of clarity):
\begin{equation}
\label{eq:covariant index up}
    \overline\nabla_\mu h^{\alpha \beta} = \de_\mu h^{\alpha \beta} + \overline{\Gamma_{\mu \lambda}^\alpha} \,h^{\lambda \beta}+ \overline{\Gamma_{\mu \lambda}^\beta} \, h^{\alpha \lambda}\,,
\end{equation}
or:
\begin{equation}
\label{eq:covariant index down}
     \overline\nabla_\mu h_{\alpha \beta} = \de_\mu h_{\alpha \beta} - \overline{\Gamma_{\mu \alpha}^\lambda}\, h_{\lambda \beta}- \overline{\Gamma_{\mu \beta}^\lambda} \,h_{\alpha \lambda}\,,
\end{equation}
where the bars over the Christoffel symbols indicate that they are computed with respect to the background metric $\bar g_{\mu \nu}$. Given that we are interested in the $(0,0)$ component of the Isaacson tensor in \eqref{eq:Isaacson tensor}, the only Christoffel symbols which are useful for our purpose are those of the form $\Gamma_{0\mu}^\nu$, which vanishes identically for $\mu=0$ or $\nu=0$. On the other hand, we have that:
\begin{equation}
\label{eq:Christoffel FLRW}
    \Gamma_{0i}^j = H\,\delta_i^j\,.
\end{equation}
Using \eqref{eq:covariant index up} and \eqref{eq:covariant index down} we get (dropping the bars):
\begin{eqnarray}
\nabla_0h^{ij} \nabla_0h_{ij} & = & \dot h^{ij} \dot h_{ij} - \Gamma^k_{0 i}\, \dot h^{i j}h_{k j} - \Gamma^k_{0 j}\, \dot h^{i j}h_{i k} \nonumber \\
& +& \Gamma^i_{0k}\, \dot h_{i j}h^{k j} + \Gamma^j_{0k}\, \dot h_{ij}h^{i k} \label{eq:nabla h nabla h} \\
& -&\Gamma_{0 k}^i \Gamma_{0 i}^\rho \, h_{\rho j} h^{k j} - \Gamma_{0k}^i \Gamma_{0 j}^r\,  h_{i r} h^{k j} \nonumber \\
& -& \Gamma_{0 k}^j \Gamma_{0 i}^r\,  h_{r j} h^{i k}- \Gamma_{0 k}^j \Gamma_{0 j}^\rho\, h_{i \rho} h^{i k}\,. 
\nonumber
\end{eqnarray}
By substituting \eqref{eq:Christoffel FLRW} in \eqref{eq:nabla h nabla h}, and using $\dot h^{ij} = a^{-4} \delta^{ik}\delta^{jl} \dot h_{kl} -4Hh^{ij}$, we find that the terms which are proportional to a single time derivative, $\dot h_{ij}$, cancel off. Substituting the covariant derivatives in \eqref{eq:rho_GW EM tensor}, we get:
\begin{equation}
\label{eq:rho_GW FLRW}
    \rho_{\rm GW} \simeq M_p^2 \left(\left\langle \dot h^{ij}\dot h_{ij} \right \rangle + 4 H^2 \left \langle h^{ij}h_{ij}\right \rangle\right).
\end{equation}
For an approximate plane wave solution:
\begin{equation}
h_{ij}(x) \simeq A\,\epsilon_{ij}\,   e^{i (\omega t- \vec k \cdot \vec x)}\,,
\end{equation}
we find:
\begin{equation}
\label{eq:rho_GW frequency}
\rho_{\rm GW} \simeq M_p^2 A^2 \left(\omega^2+4H^2\right)\,.
\end{equation}
Upon emission from the string loops, the GW background grows up to an order $1$ fraction of the total energy of the Universe, i.e. $\rho_{\rm GW} \simeq \rho_{\rm tot} = 3 M_p^2 H^2$. It follows that:
\begin{equation}
\label{eq:h=H/omega}
A \simeq \frac{H}{\omega} \left[1+\left(\frac{2 H}{\omega}\right)^2\right]^{-1/2}\,.
\end{equation}
However, we know that the peak emission frequency of the loops corresponds to $\omega \simeq 1/\ell$. Hence, in the subhorizon limit $H \ell \ll 1$, we have $H\ll\omega$, implying:
\begin{equation}
\label{eq:h=Hl}
A \simeq H /\omega \ll 1\,,
\end{equation}
which shows that perturbation theory is under control, since the condition \eqref{eq:perturbation condition} is satisfied. 

\end{appendix}

\newpage

\bibliographystyle{JHEP}
\bibliography{references}

\end{document}